\documentclass{revtex4}
\usepackage{amsmath}
\usepackage{amssymb}
\usepackage{amsfonts}
\usepackage{bm}
\usepackage[dvips]{color}
\usepackage[dvips]{graphicx}

\begin{document}

%------------------------------------------------------------------------------
% title
\title{Coarse-Graining of Microscopic Dynamics into \\
Mesoscopic Transient Potential Model}

\author{Takashi Uneyama}
\affiliation{%
%\email{uneyama@mp.pse.nagoya-u.ac.jp}
  JST, PRESTO, and Center for Computational Science, Graduate School of Engineering,
  Nagoya University, Furo-cho, Chikusa, Nagoya 464-8603,
  Japan
}%

\begin{abstract}
We show that a mesoscopic coarse-grained dynamics model
which incorporates the transient potential 
can be formally derived from an underlying microscopic
dynamics model.
As a microscopic dynamics model, we employ the overdamped Langevin equation.
By utilizing the path probability and the Onsager-Machlup type action,
we calculate the path probability for the coarse-grained mesoscopic
degrees of freedom. The action for the mesoscopic degrees of freedom
can be simplified by incorporating the transient potential.
Then the dynamic equation for the mesoscopic degrees of
freedom can be simply described by the Langevin equation with the
transient potential (LETP).
As a simple and analytically tractable approximation,
we introduce additional degrees of freedom which express the state
of the transient potential.
Then we approximately express the dynamics of the system as the
the combination of the LETP and the dynamics model for the transient potential.
The resulting dynamics model
has the same dynamical structure as the responsive particle dynamics (RaPiD)
type models [W. J. Briels, Soft Matter {\bf 5}, 4401 (2009)] and the
multi-chain slip-spring type models [T. Uneyama and Y. Masubuchi, J. Chem. Phys. {\bf 137}, 154902 (2012)].
As a demonstration, we apply our coarse-graining method with the LETP
to a single particle
dynamics in a supercooled liquid, and compare the results of the LETP
with the molecular dynamics simulations and other coarse-graining models.
\end{abstract}

\maketitle

%------------------------------------------------------------------------------
% main text
%

%------------------------------------------------------------------------------
\section{Introduction}
\label{introduction}

Soft matters such as polymers form various mesoscopic structures and
exhibit various interesting dynamics.
Coarse-grained models are useful to study the mesoscopic dynamics of
such complex systems by simulations, especially at the long time scale.
The coarse-graining reduces the
degrees of freedom of the system, and changes the characteristic time and
length scales. As a result, the computational costs required for simulations
drastically reduce. For some soft matter systems such as polymer melts,
due to their long relaxation times, we cannot study their long time relaxation
behavior without coarse-grained models\cite{MullerPlathe-2002,Padding-Briels-2011}.
Although the coarse-grained models are useful for simulations, the validity of
simulation results are not always guaranteed. This is because the
coarse-graining processes usually involve some approximations, and the validity
of coarse-grained models strongly depends on the properties of the
employed approximations. Unfortunately, the properties of approximations are
not clear in some cases. Some coarse-grained models, such as the reptation
model for entangled polymers\cite{Doi-Edwards-book}, are rather phenomenologically
proposed, and not theoretically derived from the underlying microscopic models.
For such cases, the relation between the microscopic models and
mesoscopic coarse-grained models is not clear in general.

To study
the properties of the coarse-grained models, theoretical methods based
on statistical mechanics are useful. If the target system is
not largely deviated from the equilibrium state, we can utilize the
linear nonequilibrium statistical mechanics. The dynamic equations for
coarse-grained degrees of freedom can be expressed, for example, as
the Langevin equation\cite{Itami-Sasa-2017} or
the generalized Langevin equation (GLE)\cite{Kawasaki-1973}. The transport coefficients can be
related to the correlation functions of underlying microscopic dynamics,
by the fluctuation-dissipation relation\cite{Evans-Morris-book}. The GENERIC (general equation
for nonequilibrium reversible-irreversible coupling) formalism
\cite{Grmela-Ottinger-1997,Ottinger-Grmela-1997,Espanol-2004} gives
a general form of the effective dynamic equations. The theoretical
analyses of the coarse-grained models from such view points are important to
understand them in detail. For example, the dissipative particle dynamics (DPD),
which was originally introduced phenomenologically, has been theoretically
justified by using some statistical mechanical methods\cite{Espanol-Warren-1995,Kinjo-Hyodo-2007,Espanol-2009}.

For entangled polymer melts which exhibit characteristic slow relaxation
behavior, various mesoscale phenomenological models have been proposed and
utilized\cite{Doi-Edwards-book}. Among them,
some recently proposed models have interesting
theoretical structures, from the view point of statistical mechanics.
Kindt and Briels proposed the responsive particle dynamics (RaPiD) model\cite{Kindt-Briels-2007,Briels-2009,Briels-2015},
in which a single polymer chain is expressed as a single
coarse-grained particle. In the RaPiD model, the number of entanglements
between different polymer chains is employed as a fluctuating dynamical
variable. The system is expressed by the particle positions
and the numbers of entanglements between particles. Then the
dynamics is described by the dynamic equations for the particles and
the numbers of entanglements. Chappa et al\cite{Chappa-Morse-Zippelius-Muller-2012}, and Uneyama and Masubuchi\cite{Uneyama-Masubuchi-2012} proposed
the multi-chain slip-spring (MCSS) model. In the MCSS model, polymer chains
are modeled as Rouse chains, and chains are connected by so-called
slip springs. The slip springs move along the chains, and are dynamically
reconstructed at chain ends. In the MCSS model, the system is 
expressed by the positions of beads which construct polymer chains,
and the states of slip-springs. The dynamics is described by the dynamic
equation for beads and some stochastic transition rules for slip-springs.

The RaPiD and MCSS models have similar theoretical structures, and in fact,
they can be unified\cite{Uneyama-2019}.
The important point is that both the RaPiD and MCSS models employ some
extra degrees of freedom (the numbers of entanglements or the slip spring
states), in addition to the usual coarse-grained degrees of freedom (the positions of
centers of mass or beads). 
If the system obeys the GLE, the state of the target system is
fully described by the coarse-grained degrees of freedom. We may
interpret that the thermodynamic state is uniquely determined by
the coarse-grained degrees of freedom. In this sense, we may call
the coarse-grained degrees of freeedom as the thermodynamic degrees of freedom.
(The memory kernel does not affect the thermodynamic state and thus
is qualitatively different from the thermodynamic degrees of freedom.)
In the RaPiD and MCSS models, in contrast, the thermodynamic potential
explicitly depends both on the coarse-grained and extra degrees of freedom.
In this work, we may call such extra degrees of freedom as
the ``pseudo thermodynamic degrees of freedom''.
The pseudo thermodynamic degrees of freedom dynamically modulate the effective
potentials for the normal degrees of freedom. This dynamic modulation
is realized through interaction potentials which are called the
``transient potentials''\cite{Briels-2009,Briels-2015}.
The success of the RaPiD and MCSS models 
leads us to an idea to generalize these models.
If we can construct a general method
which employs the transient potential and pseudo thermodynamic degrees of freedom,
it will provide various mesoscopic coarse-grained dynamic equations for soft matter systems.

In this work, we show that we can actually
construct a mesoscopic coarse-grained model
with the transient potential, starting from the underlying microscopic dynamics model.
In general, we cannot obtain the dynamic equation for the transient potential
in an explicit form. We propose a simple dynamics model for the transient
potential by using the pseudo thermodynamic degrees of freedom.
We also propose some formal expressions for the dynamics of the transient potential.
We show that,
under some assumptions, we can derive the dynamic equation models which
are consistent with the RaPiD and MCSS models.
To study properties of our theoretical method in detail, we compare our
method with the GLE and the Langevin equation
with the fluctuating diffusivity. Also, we apply our model
and the GLE to the dynamics of a single tagged particle in a supercooled liquid,
and consider whether these coarse-graining methods can reasonably describe
the dynamics or not.

%------------------------------------------------------------------------------
\section{Theory}
\label{theory}

\subsection{Microscopic Model}
\label{microscopic_model}

%In this work, we employ a system which obeys an overdamped Langevin equation as
%the microscopic model.
When we consider the coarse-graining, the Hamilton's canonical equations are
employed as microscopic models in most cases\cite{Kawasaki-1973,Dengler-2016}.
However, for soft matters such as polymers, the overdamped Langevin equations are reasonably
utilized as the microscopic molecular models\cite{Doi-Edwards-book}.
In addition, by applying the standard coarse-graining procedure, one can obtain
a Langevin equation from the Hamilton's canonical equations. Therefore,
in this work, we employ an overdamped Langevin equation as the microscopic model.
We consider the microscopic model which consists of $N$ particles in a three dimensional space,
and we describe the position of the $i$-th particle as $\bm{r}_{i}$.
We employ the following Langevin equation as the microscopic dynamic equation for the $i$-th particle:
\begin{equation}
 \label{langevin_equation_microscopic}
 \frac{d\bm{r}_{i}(t)}{dt} = - \sum_{j} \bm{L}_{ij} \cdot \frac{\partial U(\lbrace \bm{r}_{i}(t)\rbrace)}{\partial \bm{r}_{i}(t)}
  + \sum_{j} \sqrt{2 k_{B} T} \bm{B}_{ij} \cdot \bm{w}_{j}(t) ,
\end{equation}
where $\bm{L}_{ij}$ is the mobility tensor, $U(\lbrace \bm{r}_{i} \rbrace)$ is the 
interaction potential energy, $\bm{B}_{ij}$ is the noise coefficient tensor which satisfies
$\bm{L}_{ij} = \sum_{k} \bm{B}_{ik} \cdot \bm{B}_{jk}^{\mathrm{T}}$  (the superscript ``$\mathrm{T}$'' represents the transpose), $k_{B}$ is the Boltzmann constant, $T$ is the temperature,
and $\bm{w}_{i}(t)$ is the Gaussian white noise. From the Onsager's reciprocal theorem,
$\bm{L}_{ij}$ is a symmetric tensor.
The noise $\bm{w}_{i}(t)$ should satisfy the following fluctuation-dissipation relation:
\begin{equation}
 \label{fluctuation_dissipation_relation_microscopic}
 \langle \bm{w}_{i}(t) \rangle = 0, \qquad
  \langle \bm{w}_{i}(t) \bm{w}_{j}(t') \rangle = \bm{1} \delta_{ij} \delta(t - t'),
\end{equation}
where $\langle \dots \rangle$ is the statistical average and $\bm{1}$ is the unit tensor.
Since eq~\eqref{langevin_equation_microscopic} is a stochastic differential equation,
we should specify the interpretation of the stochastic term\cite{Gardiner-book}. We employ the Ito
interpretation in this work. (One can employ the Stratonovich interpretation instead.
In that case, we convert the Stratonovich type equation to the Ito type equation\cite{Gardiner-book}.
The result is the same in the current case.)

For the sake of simplicity, we introduce a short-hand notation for the positions as
$\bm{R} \equiv [r_{1x},r_{1y},r_{1z},r_{2x},\dots,r_{Nz}]^{\mathrm{T}}$.
The vector $\bm{R}$ can be interpreted as a $3N$-dimensional vector.
We describe the mobility tensor, the noise coefficient tensor, and the Gaussian white noise in a similar way.
For the sake of simplicity, we also employ the short-hand notation for the
noise coefficient tensor, $\bm{B} = \bm{L}^{1/2}$. (Here, $\bm{L}^{1/2}$ represents the matrix
square root which satisfies $\bm{L}^{1/2} \cdot (\bm{L}^{1/2})^{\mathrm{T}} = \bm{L}$.)
Then, eq~\eqref{langevin_equation_microscopic} can be rewritten as
\begin{equation}
 \label{langevin_equation_microscopic_modified}
 \frac{d\bm{R}(t)}{dt} = - \bm{L} \cdot \frac{\partial U(\bm{R}(t))}{\partial \bm{R}(t)}
  + \sqrt{2 k_{B} T} \bm{L}^{1/2} \cdot \bm{w}(t) ,
\end{equation}
and eq~\eqref{fluctuation_dissipation_relation_microscopic} can be rewritten as
\begin{equation}
 \label{fluctuation_dissipation_relation_microscopic_modified}
 \langle \bm{w}(t) \rangle = 0, \qquad
  \langle \bm{w}(t) \bm{w}(t') \rangle = \bm{1} \delta(t - t') .
\end{equation}
In what follows, we use eq~\eqref{langevin_equation_microscopic_modified} 
as the microscopic dynamic equation. The equilibrium probability distribution for
the position $\bm{R}$ is simply given as the Boltzmann distribution:
\begin{equation}
 P_{\text{eq}}(\bm{R}) = \frac{1}{\mathcal{Z}}
  \exp[ - U(\bm{R}) / k_{B} T] ,
\end{equation}
where $\mathcal{Z}$ is the partition function:
\begin{equation}
 \mathcal{Z} \equiv \int d\bm{R} \, \exp[ - U(\bm{R}) / k_{B} T] .
\end{equation}
For simplicity, we have
assumed that all the particles in the system are distinguishable and
ignored the Gibbs factor.

The probability (of the realization) for the Gaussian white noise which satisfies
eq~\eqref{fluctuation_dissipation_relation_microscopic_modified} is given as\cite{Kleinert-book}
\begin{equation}
 \label{path_probability_white_noise_microscopic}
 \mathcal{P}[\bm{w}] = \mathcal{N}^{(\bm{w})} \exp\left[ - \frac{1}{2} \int dt \, \bm{w}^{2}(t) \right] ,
\end{equation}
where $\mathcal{N}^{(\bm{w})}$ is the normalization factor.
Eq~\eqref{path_probability_white_noise_microscopic}
can be interpreted as the probability of a specific path, and thus we may
call it as the path probability. The normalization factor should be 
determined so that the functional integral (path integral) over $\bm{w}$ becomes unity:
$\int \mathcal{D}\bm{w} \, \mathcal{P}[\bm{w}] = 1 $.
(In this work, however, the normalization factor itself does not become
important and thus we do not consider it in detail.)
By combining eqs~\eqref{langevin_equation_microscopic_modified} and
\eqref{path_probability_white_noise_microscopic}, the path probability for
$\bm{R}(t)$ is given as
\begin{equation}
 \label{path_probability_position_microscopic}
 \mathcal{P}[\bm{R}] = \mathcal{N}^{(\bm{R})}
  \exp\left[ - \mathcal{S}[\bm{R}] \right] ,
\end{equation}
\begin{equation}
 \label{onsager_machlup_action_microscopic}
 \mathcal{S}[\bm{R}]
  \equiv \frac{1}{2 k_{B} T} \int dt \,
  G\left(\frac{d\bm{R}(t)}{dt} + \bm{L} \cdot \frac{\partial U(\bm{R}(t))}{\partial \bm{R}(t)}; \bm{L}
      \right) ,
\end{equation}
\begin{equation}
 \label{gaussian_weight_function}
 G(\bm{x},\bm{C})
  \equiv \frac{1}{2} \bm{x}^{\mathrm{T}} \cdot \bm{C}^{-1} \cdot \bm{x},
\end{equation}
where $\mathcal{N}^{(\bm{R})}$ is the normalization factor (and is
generally different from $\mathcal{N}^{(\bm{w})}$, due to the Jacobian for the variable transform), $\mathcal{S}[\bm{R}]$ is the action which gives the statistical weight
for a specific path (the Onsager-Machlup action)\cite{Onsager-Machlup-1953,Machlup-Onsager-1953}.
In what follows, we express normalization factors for the path probabilities
by $\mathcal{N}^{(\dotsb)}$ in a similar way.
Eq~\eqref{gaussian_weight_function} represents the Gaussian weight for a vector $\bm{x}$ and a covariance tensor $\bm{C}$.
The covariance tensor $\bm{C}$ is a second rank symmetric positive definite tensor and
$\bm{C}^{-1}$ is its inverse:
$\bm{C} \cdot \bm{C}^{-1} = \bm{1}$.
All the information on the microscopic
dynamics is given by the path probability \eqref{path_probability_position_microscopic}.

Before we consider the coarse-graining of the microscopic dynamic equation,
here we briefly comment about the mobility model. In eq~\eqref{langevin_equation_microscopic_modified},
the mobility tensor $\bm{L}$ is assumed to be independent of the position vector $\bm{R}(t)$.
Such a situation is realized, for example, if we consider the situation where each particles feel the
friction independently. The noise term is statistically independent of
$\bm{R}(t)$ (the additive noise), and the analyses can be simplified.
However, in general, the mobility tensor can depend on $\bm{R}(t)$, such
as the case of the systems with the hydrodynamic interaction.
If the mobility tensor depends on $\bm{R}(t)$, then the noise coefficient
tensor $\bm{L}^{1/2}$ also depends on $\bm{R}(t)$. In such a case, the noise term becomes
the multiplicative noise.
The extension of our theory to the multiplicative noise is possible but complicated.
(We show the extension in Appendix~\ref{multipliticative_noise_and_nonlinear_variable_transform}.)
Thus here we limit ourselves to the case of the additive noise.

\subsection{Coarse-Graining}
\label{coarse_graining}

What we want to obtain here is the effective dynamic equation for some mesoscopic degrees of
freedom. We limit ourselves that the mesoscopic degrees of freedom which
can be given as the linear combinations the microscopic position, $\bm{R}(t)$.
(The nonlinear variable transform can be employed but the calculation becomes
complicated. We show the extension of the theory to the nonlinear variable transform in Appendix~\ref{multipliticative_noise_and_nonlinear_variable_transform}.)
For example, the centers of mass of molecules and the end-to-end vectors of polymers
can be expressed as the linear combinations. We describe the $i$-th mesoscopic degrees 
of freedom as $Q_{i}$, and assume that there are $M$ mesoscopic variables.
(The number of mesoscopic variables $M$ is generally much smaller than the
number of microscopic degrees of freedom, $3 N$.)
Then, without loss of generality, we can transform the microscopic degrees of
freedom $\bm{R}$ as
\begin{equation}
 \label{variable_transform_microscopic_to_mesoscopic}
  \bm{X} \equiv
 \begin{bmatrix}
  \bm{Q} \\
  \bm{\theta}
 \end{bmatrix}
 = \bm{V} \cdot \bm{R} ,
\end{equation}
where $\bm{Q} = [Q_{1},Q_{2},\dots,Q_{M}]^{\mathrm{T}}$ (an $M$-dimensional vector),
$\bm{\theta}$ is a $(3N - M)$-dimensional vector, and $\bm{V}$ is a transformation matrix (of which dimension is $3 N \times 3 N$).
We can take $\bm{\theta}$ so that the transformation matrix is invertible.
Then we can express $\bm{R}$ as follows, by inverting eq~\eqref{variable_transform_microscopic_to_mesoscopic}:
\begin{equation}
 \label{variable_transform_mesoscopic_to_microscopic}
 \bm{R} = \bm{V}^{-1} \cdot \begin{bmatrix}
  \bm{Q} \\
  \bm{\theta}
 \end{bmatrix} = \bm{V}^{-1} \bm{X}.
\end{equation}
From eqs \eqref{variable_transform_microscopic_to_mesoscopic}, a function
of $\bm{R}$ such as the potential energy $U$ can be interpreted as
a function of $\bm{X}$ (or, equivalently, a function of $\bm{Q}$ and $\bm{\theta}$).

We rewrite eqs~\eqref{path_probability_position_microscopic} and \eqref{onsager_machlup_action_microscopic}
as functionals of $\bm{Q}$ and $\bm{\theta}$:
\begin{equation}
 \label{path_probability_position_mesoscopic}
 \mathcal{P}[\bm{Q},\bm{\theta}] = \mathcal{N}^{(\bm{Q},\bm{\theta})}
  \exp\left[ - \mathcal{S}[\bm{Q},\bm{\theta}] \right] ,
\end{equation}
\begin{equation}
 \label{onsager_machlup_action_mesoscopic}
 \mathcal{S}[\bm{Q},\bm{\theta}]
  = \frac{1}{2 k_{B} T}
 \int dt \, 
 G\left( \frac{d \bm{X}(t)}{dt} + \bm{L}' \cdot \frac{\partial U(\bm{X}(t))}{\partial \bm{X}(t)} ; \bm{L}' \right) ,
\end{equation}
where $\bm{L}' \equiv \bm{V} \cdot \bm{L} \cdot \bm{V}^{-1}$ is the mobility tensor for $\bm{X}$.
The path probability for the mesoscopic degrees of freedom can be obtained by
eliminating the variable $\bm{\theta}$:
\begin{equation}
 \label{effective_path_probability_position_mesoscopic}
 \mathcal{P}[\bm{Q}] = \int \mathcal{D}\bm{\theta} \, \mathcal{P}[\bm{Q},\bm{\theta}] .
\end{equation}
Unfortunately, $\bm{Q}$ and $\bm{\theta}$ are coupled in a complicated way.
In general, we cannot evaluate eq~\eqref{effective_path_probability_position_mesoscopic} analytically.
We need to introduce some approximations to proceed the calculation.

Here, we recall that the vector $\bm{\theta}$ can be arbitrarily chosen as long as
$\bm{V}$ is invertible. Because we are interested only on the mesoscopic
variable $\bm{Q}$, the choice of $\bm{\theta}$ is still rather arbitrarily at
this stage. We choose $\bm{\theta}$ so that the action
becomes a simple form. We employ $\bm{\theta}$ which gives the following mobility tensor
\begin{equation}
 \bm{L}' =
  \begin{bmatrix}
   \bm{\Lambda} & 0 \\
   0 & \bm{M}
  \end{bmatrix} ,
\end{equation}
where $\bm{\Lambda}$ and $\bm{M}$ are the mobility tensors 
for $\bm{Q}$ and $\bm{\theta}$, respectively.
(The dimensions of $\bm{\Lambda}$ and $\bm{M}$ are $M \times M$ and $(3 N - M) \times (3 N - M)$, respectively.)
In other words, we employ $\bm{\theta}$ which is  $\bm{L}'$-orthogonal to $\bm{Q}$:
\begin{equation}
 \begin{bmatrix}
  \bm{Q}^{\mathrm{T}} & 0
 \end{bmatrix} \cdot  \bm{L}' \cdot
 \begin{bmatrix}
  0 \\
  \bm{\theta}
 \end{bmatrix} = 0.
\end{equation}
%$\bm{Q}^{\mathrm{T}} \cdot \bm{L}' \cdot \bm{\theta} = 0$.
With this specific choice of $\bm{\theta}$, we can further rewrite eq~\eqref{onsager_machlup_action_mesoscopic} as
\begin{equation}
 \label{onsager_machlup_action_mesoscopic_modified}
   \mathcal{S}[\bm{Q},\bm{\theta}]
    = 
   \mathcal{S}^{(\bm{Q})}[\bm{Q}|\bm{\theta}]
   + \mathcal{S}^{(\bm{\theta})}[\bm{\theta}|\bm{Q}] ,
\end{equation}
\begin{equation}
 \label{onsager_machlup_action_mesoscopic_position}
   \mathcal{S}^{(\bm{Q})}[\bm{Q}|\bm{\theta}]
    \equiv \frac{1}{2 k_{B} T} \int dt \,
    \left[  G\left(\frac{d \bm{Q}(t)}{dt} + \bm{\Lambda} \cdot \frac{\partial U(\bm{Q}(t),\bm{\theta}(t))}{\partial \bm{Q}(t)}
   ;  \bm{\Lambda}
      \right) \right] ,
\end{equation}
\begin{equation}
 \label{onsager_machlup_action_mesoscopic_remaining}
   \mathcal{S}^{(\bm{\theta})}[\bm{\theta}|\bm{Q}]
    \equiv \frac{1}{2 k_{B} T} \int dt \,
   \left[ 
   G\left(\frac{d \bm{\theta}}{dt} + \bm{M} \cdot \frac{\partial U(\bm{Q}(t),\bm{\theta}(t))}{\partial \bm{\theta}(t)}
      ; \bm{M} \right) 
   \right] .
\end{equation}
In eq~\eqref{onsager_machlup_action_mesoscopic_modified}, the Gaussian weight
factor is split into two contributions (eqs~\eqref{onsager_machlup_action_mesoscopic_position}
and \eqref{onsager_machlup_action_mesoscopic_remaining}), unlike that in eq~\eqref{onsager_machlup_action_mesoscopic}.
However, it should be noticed that
two split weight factors are coupled through the interaction potential $U(\bm{Q},\bm{\theta})$.
Thus we cannot simply eliminate the degrees of freedom $\bm{\theta}$ by
performing the functional integral over $\bm{\theta}$.

The Onsager-Machlup action \eqref{onsager_machlup_action_mesoscopic_modified}
gives the statistical weight for a certain path\cite{Martin-Siggia-Rose-1973}. This is in analogy to
the free energy functional in the field theory\cite{Onuki-book}; the free energy functional
gives the statistical weight for a certain field. In the field theory,
we often introduce some auxiliary fields to obtain the approximate expression
for the free energy. We expect that the action can be approximated in
a similar way. We introduce a transient potential as an auxiliary variable.
We interpret the potential at time $t$, $U(\bm{Q}(t),\bm{\theta}(t))$,
as a transient potential $\Phi(\bm{Q}(t),t)$. This transient potential $\Phi$ is 
a function of $\bm{Q}$ and $t$, and is independent of $\bm{\theta}$.
Following the standard procedure in the field theory
\cite{Muller-Schmid-2005,Kawakatsu-book}, we use the following identity for the delta functional:
\begin{equation}
 \label{delta_functional_identity}
 1 = \int \mathcal{D}\Phi \, 
  \delta\left[ \Phi(\tilde{\bm{q}},t) - U(\tilde{\bm{q}},\bm{\theta}(t)) \right] .
\end{equation}
Here, $\tilde{\bm{q}}$ represents the dummy variable which
has the same dimension as $\bm{Q}$.
By inserting eq~\eqref{delta_functional_identity} into
eq~\eqref{effective_path_probability_position_mesoscopic}, we have
\begin{equation}
 \label{effective_path_probability_position_mesoscopic_modified}
  \begin{split}
   \mathcal{P}[\bm{Q}]
   & = \int \mathcal{D}\bm{\theta}\mathcal{D}\Phi \, \delta\left[ \Phi(\tilde{\bm{q}},t) - U(\tilde{\bm{q}},\bm{\theta}(t)) \right] \mathcal{N}^{(\bm{Q},\bm{\theta})}
  \exp\left[ - \mathcal{S}[\bm{Q},\bm{\theta}] \right] \\
   & = \int \mathcal{D}\Phi \, \mathcal{N}^{(\bm{Q},\Phi)}
  \exp\left[ - \tilde{\mathcal{S}}^{(\bm{Q})}[\bm{Q}|\Phi]
   \right] \tilde{\mathcal{P}}^{(\Phi)}[\Phi|\bm{Q}] ,
  \end{split}
\end{equation}
\begin{equation}
 \label{onsager_machlup_action_mesoscopic_modified2}
   \tilde{\mathcal{S}}^{(\bm{Q})}[\bm{Q}|\Phi]
    \equiv \frac{1}{2 k_{B} T} \int dt \,
  G\left(\frac{d \bm{Q}(t)}{dt} + \bm{\Lambda} \cdot \frac{\partial \Phi(\bm{Q}(t),t)}{\partial \bm{Q}(t)}
   ;  \bm{\Lambda}
      \right) ,
\end{equation}
\begin{equation}
 \label{effective_path_probability_remaining}
  \begin{split}
   \tilde{\mathcal{P}}^{(\Phi)}[\Phi|\bm{Q}]
   & \equiv\int \mathcal{D}\bm{\theta} \, \delta\left[ \Phi(\tilde{\bm{q}},t) - U(\tilde{\bm{q}},\bm{\theta}(t)) \right] \\
   & \qquad \times \exp\left[ - \frac{1}{2 k_{B} T} \int dt \,
   G\left(\frac{d \bm{\theta}}{dt} + \bm{M} \cdot \frac{\partial U(\bm{Q}(t),\bm{\theta}(t))}{\partial \bm{\theta}(t)}
      ; \bm{M} \right) 
  \right] .
  \end{split}
\end{equation}
$\tilde{\mathcal{S}}^{(\bm{Q})}[\bm{Q}|\Phi]$ (eq~\eqref{onsager_machlup_action_mesoscopic_modified2})
can be interpreted as the action for $\bm{Q}$
under a given $\Phi$. Similarly, $\tilde{\mathcal{P}}^{(\Phi)}[\Phi|\bm{Q}]$
(eq~\eqref{effective_path_probability_remaining}) can be interpreted
as the path probability for $\Phi$ under a given $\bm{Q}$. For convenience,
we introduce the action for $\Phi$ and rewrite eq~\eqref{effective_path_probability_position_mesoscopic_modified} as
\begin{equation}
 \label{effective_path_probability_position_mesoscopic_modified2}
   \mathcal{P}[\bm{Q}]
    = \int \mathcal{D}\Phi \, \mathcal{N}^{(\bm{Q},\Phi)}
  \exp\left[ - \tilde{\mathcal{S}}^{(\bm{Q})}[\bm{Q}|\Phi]
       - \tilde{\mathcal{S}}^{(\Phi)}[\Phi|\bm{Q}]
   \right] ,
\end{equation}
\begin{equation}
 \label{onsager_machlup_action_remaining}
   \tilde{\mathcal{S}}^{(\Phi)}[\Phi|\bm{Q}] \equiv - \ln \tilde{\mathcal{P}}^{(\Phi)}[\Phi|\bm{Q}] .
\end{equation}

So far, we have not introduced any approximations for the Onsager-Machlup action.
Thus eq~\eqref{effective_path_probability_position_mesoscopic_modified2} is
exactly equivalent to eq~\eqref{effective_path_probability_position_mesoscopic}.
Of course, eq~\eqref{effective_path_probability_position_mesoscopic_modified2} is
just a formal expression and we have no simple analytic expression for the action
$\tilde{\mathcal{S}}^{(\Phi)}$. Nonetheless eq~\eqref{effective_path_probability_position_mesoscopic_modified2}
is useful for the coarse-graining. Eq~\eqref{effective_path_probability_position_mesoscopic_modified2} implies that, the transient potential $\Phi$ can be employed
as additional degrees of freedom of the mesoscopic system.
Instead of the path probability for $\bm{Q}$ (as eq~\eqref{effective_path_probability_position_mesoscopic}), here we consider the
path probability for $\bm{Q}$ and $\Phi$:
\begin{equation}
 \label{effective_path_probability_position_transient_potential}
   \mathcal{P}[\bm{Q},\Phi]
    \equiv  \mathcal{N}^{(\bm{Q},\Phi)}
  \exp\left[ - \tilde{\mathcal{S}}^{(\bm{Q})}[\bm{Q}|\Phi]
       - \tilde{\mathcal{S}}^{(\Phi)}[\Phi|\bm{Q}]
   \right] .
\end{equation}
Clearly, we have $\int \mathcal{D}\Phi \mathcal{P}[\bm{Q},\Phi] = \mathcal{P}[\bm{Q}]$.
Thus, if we eliminate the transient potential from eq \eqref{effective_path_probability_position_transient_potential},
we recover the path probability for $\bm{Q}$. Now we have two actions in
eq~\eqref{effective_path_probability_position_transient_potential}.
The action for $\bm{Q}$, $\tilde{\mathcal{S}}^{(\bm{Q})}$ (eq~\eqref{onsager_machlup_action_mesoscopic_modified2}),
is simple and we need no further manipulation for it (as long as $\Phi$ is given).
The Langevin equation which corresponds to the action \eqref{onsager_machlup_action_mesoscopic_modified2} is
\begin{equation}
 \label{langevin_equation_mesoscopic_transient_potential}
 \frac{d\bm{Q}(t)}{dt} = - \bm{\Lambda} \cdot \frac{\partial \Phi(\bm{Q}(t),t)}{\partial \bm{Q}(t)} + \sqrt{2 k_{B} T} \bm{\Lambda}^{1/2} \cdot \bm{W}(t),
\end{equation}
where $\bm{W}(t)$ is 
the $M$-dimensional Gaussian white noise vector. The noise $\bm{W}$ satisfies
\begin{equation}
 \label{fluctuation_dissipation_mesoscopic_transient_potential}
 \langle \bm{W}(t) \rangle = 0, \qquad
  \langle \bm{W}(t) \bm{W}(t') \rangle = \bm{1} \delta(t - t').
\end{equation}
On the other hand, the action for $\Phi$, which is given by
eqs \eqref{effective_path_probability_remaining} and \eqref{onsager_machlup_action_remaining},
is not simple. From eq~\eqref{onsager_machlup_action_remaining},
the transient potential $\Phi$ obeys a stochastic time evolution equation
(such as the Langevin equation), and this equation depends on $\bm{Q}$.
We need to approximate it by a simple and tractable form, in order to obtain a dynamic
equation model which is suitable for numerical simulations and theoretical analyses.
Once we have the (approximate) dynamic equation for the transient potential,
we can combine it with eq~\eqref{langevin_equation_mesoscopic_transient_potential} to
describe the dynamics of the mesoscopic degrees of freedom.
Therefore, we find that the dynamics for the mesoscopic degrees of freedom
is described by the Langevin equation with the transient potential (LETP).
Eqs~\eqref{effective_path_probability_position_transient_potential}
and \eqref{langevin_equation_mesoscopic_transient_potential}
formally justify the dynamics models with transient potentials,
which were originally proposed as phenomenological models.

We can derive a similar Langevin equation for the cases where
the noise is multiplicative and/or the variable transform is nonlinear.
In general, the mobility tensor $\bm{\Lambda}$ becomes time-dependent and
fluctuating quantity, just like the transient potential $\Phi$.
The detailed calculations are shown in Appendix~\ref{multipliticative_noise_and_nonlinear_variable_transform}.
In what follows, for the sake of simplicity, we consider
only the case of the additive noise and the linear variable transform.

\subsection{Dynamics Model for Transient Potential}
\label{dynamics_model_for_transient_potential}

We should notice that our procedure in Sec.~\ref{coarse_graining} does not
give the information
on the dynamics of the transient potential. The derivation above is 
formal and one may criticize that it does not fully justify the LETP and
thus cannot be accepted as a concrete derivation.
Such a criticism is partly true.
However, generally we cannot obtain the ``exact'' dynamic equations for coarse-grained
systems. We need to employ some approximations for the full
dynamics model to obtain a coarse-grained model, but approximations may not be
fully justified and are rather empirical.
In this subsection, we consider some methods to determine the effective
dynamics model for the transient potential. We cannot determine the
dynamics model uniquely, but we show that we can construct physically
reasonable models under given approximations.

We start from a rather formal expression. The dynamics of the transient
potential can be formally determined by the action $\tilde{\mathcal{S}}^{(\Phi)}[\Phi|\bm{Q}]$.
From eq~\eqref{effective_path_probability_position_transient_potential}, we have
\begin{equation}
 \label{effective_path_probability_remaining_modified}
  \tilde{\mathcal{S}}^{(\Phi)}[\Phi|\bm{Q}]  =
  - \ln \mathcal{P}[\bm{Q},\Phi]
  - \tilde{\mathcal{S}}^{(\bm{Q})}[\bm{Q}|\Phi]
  + (\text{const.}).
\end{equation}
The explicit form of the action $\tilde{\mathcal{S}}^{(\bm{Q})}[\bm{Q}|\Phi]$ is given by eq~\eqref{onsager_machlup_action_mesoscopic_modified2}.
Also, the path probability $\mathcal{P}[\bm{Q},\Phi]$ can be obtained as the
ensemble average as:
\begin{equation}
 \label{effective_path_probability_position_transient_potential_direct}
 \mathcal{P}[\bm{Q},\Phi] =
  \left\langle \delta(\bm{Q}(t) - \bm{Q}(\hat{\bm{R}}(t))) \, \delta[\Phi(\tilde{\bm{q}},t) - U(\tilde{\bm{q}},\bm{\theta}(\hat{\bm{R}}(t))) ]\right\rangle^{(\hat{\bm{R}})}.
\end{equation}
Here, $\hat{\bm{R}}$ represents a trajectory (or a path)
directly generated by the Langevin equation \eqref{langevin_equation_microscopic_modified},
and the statistical average $\langle \dots \rangle^{(\hat{\bm{R}})}$ is taken for realizations of $\hat{\bm{R}}$.
Therefore, in principle, we can construct the action $\tilde{\mathcal{S}}^{(\Phi)}[\Phi|\bm{Q}]$ from
the path probability calculated by the direct microscopic simulations.
Of course, the calculation of the path probability by
eq~\eqref{effective_path_probability_position_transient_potential_direct}
is practically impossible since the path probability is the joint distribution functional
for the path and potential function in very high dimensions.

We consider to construct a dynamics model which can be used for practical
simulations and analyses, by introducing some approximations.
Even if the resulting dynamics model for the transient potential
is not exact, the model which mimics the exact dynamics and
gives physically reasonable results would be still useful.
We assume that the transient potential
can be approximately expressed as a function of a $Z$-dimensional
auxiliary variable $\bm{A}(t)$:
\begin{equation}
 \label{transient_potential_approximation}
 \Phi(\tilde{\bm{q}},t) \approx \check{\Phi}(\tilde{\bm{q}},\bm{A}(t)).
\end{equation}
The auxiliary variable $\bm{A}(t)$ should be chosen so that it gives
a reasonable approximation for the dynamics of the transient potential.
The dimension $Z$ should be sufficiently smaller than the dimension of $\bm{\theta}$, $Z \ll (3 N - M)$.
$\bm{A}(t)$ does not need to have the expression in terms of $\bm{\theta}(t)$.
From eq~\eqref{transient_potential_approximation},
$\bm{A}(t)$ can be interpreted as a sort of the state
of the transient potential. Then we expect that it behaves in a similar
way to the coarse-grained variable $\bm{Q}(t)$.
We further assume that, in equilibrium,
the joint probability of $\bm{Q}$ and $\bm{A}$ should be expressed as
\begin{equation}
 \label{equilibrium_distribution_with_auxiliary_variable}
 P_{\text{eq}}(\bm{Q},\bm{A}) = \frac{1}{\check{\mathcal{Z}}}
  \exp[- \check{\Phi}(\bm{Q},\bm{A}) / k_{B} T] ,
\end{equation}
where $\check{\mathcal{Z}}$ is the effective partition function:
\begin{equation}
 \check{\mathcal{Z}} = \int d\bm{Q} d\bm{A} \,
  \exp[- \check{\Phi}(\bm{Q},\bm{A}) / k_{B} T] .
\end{equation}
Eq~\eqref{equilibrium_distribution_with_auxiliary_variable} is the same
form as the usual partition function. Thus our assumption for $\bm{A}$ is
that, it behaves as usual degrees of freedom. The thermodynamic
state of the coarse-grained model is usually determined by the coarse-grained
variable $\bm{Q}$. (As we stated, we may call such coarse-grained variables
as the thermodynamic degrees of freedom, in this work.)
In a similar way, we assume that the thermodynamic
state can be now determined by $\bm{Q}$ and $\bm{A}$. Therefore we may
call the auxiliary variable $\bm{A}$ as pseudo thermodynamic degrees of
freedom.

Since the pseudo thermodynamic degrees of freedom were introduced 
to approximately describe the dynamics for the transient potential, they should never affect the equilibrium
statistics of the mesoscopic degrees of freedom. Thus we require
\begin{equation}
 P_{\text{eq}}(\bm{Q}) = \int d\bm{A} \, P_{\text{eq}}(\bm{Q},\bm{A}) ,
\end{equation}
or, equivalently,
\begin{equation}
 \label{free_energy_and_transient_potential}
 \exp[ -\mathcal{F}(\bm{Q}) / k_{B} T] = \int d\bm{A} \,
  \exp[ - \check{\Phi}(\bm{Q},\bm{A}) / k_{B} T] ,
\end{equation}
where $\mathcal{F}(\bm{Q})$ is the free energy for the mesoscopic degrees of freedom $\bm{Q}$:
\begin{equation}
 \mathcal{F}(\bm{Q}) \equiv - k_{B} T \ln \int d\bm{\theta} \exp[-U(\bm{Q}, \bm{\theta}) / k_{B} T].
\end{equation}
From eq~\eqref{free_energy_and_transient_potential}, we can relate the forces by the transient potential and the free energy as
\begin{equation}
 \label{average_thermodynamic_force_transient_potential}
 \frac{\partial \mathcal{F}(\bm{Q})}{\partial \bm{Q}}
  = \int d\bm{A} \, \frac{\partial \check{\Phi}(\bm{Q},\bm{A})}{\partial \bm{Q}} P_{\text{eq}}(\bm{Q},\bm{A}) .
\end{equation}
The physical meaning of eq~\eqref{average_thermodynamic_force_transient_potential} is
clear. If we average the thermodynamic force by the transient
potential over the pseudo thermodynamic degrees of freedom, we just have the thermodynamic force by the free energy $\mathcal{F}$.
Therefore, if the pseudo thermodynamic degrees of freedom relax much rapidly compared with
the mesoscopic degrees of freedom, we just have a usual Langevin equation.

We want the dynamics model for $\bm{A}(t)$
to be simple and free from the memory kernel.
We assume that $\bm{A}(t)$ obeys a Markovian stochastic process.
We express the probability distribution of $\bm{Q}$ and $\bm{A}$ at time
$t$ as $P(\bm{Q},\bm{A};t)$. For a Markovian process, the time evolution of $P(\bm{Q},\bm{A},t)$ can
be formally expressed as follows.
\begin{align}
 \label{master_equation_approx}
 \frac{\partial P(\bm{Q},\bm{A};t)}{\partial t}
 & = [\mathcal{L}^{(\bm{Q})} + \mathcal{L}^{(\bm{A})}] P(\bm{Q},\bm{A};t), \\
 \label{fokker_plack_operator}
 \mathcal{L}^{(\bm{Q})} P(\bm{Q},\bm{A})
 & = \frac{\partial}{\partial \bm{Q}} \cdot \bm{\Lambda} \cdot
 \left[ \frac{\partial \check{\Phi}(\bm{Q},\bm{A})}{\partial \bm{Q}} P(\bm{Q},\bm{A}) 
 + k_{B} T \frac{\partial P(\bm{Q},\bm{A}) }{\partial \bm{Q}} \right], \\
 \mathcal{L}^{(\bm{A})} P(\bm{Q},\bm{A})
 \label{transition_rate_matrix_approx}
 & = \int d\bm{A}' \, [\check{\Omega}(\bm{A}|\bm{A}',\bm{Q}) P(\bm{Q},\bm{A}')
 - \check{\Omega}(\bm{A}'|\bm{A},\bm{Q}) P(\bm{Q},\bm{A}) ].
\end{align}
Eq~\eqref{fokker_plack_operator} is derived from the Langevin equation \eqref{langevin_equation_mesoscopic_transient_potential}
with the approximate transient potential \eqref{transient_potential_approximation}.
$\check{\Omega}(\bm{A}'|\bm{A},\bm{Q})$ is the transition rate from
$\bm{A}$ to $\bm{A}'$, and it
should satisfy the detailed-balance condition:
\begin{equation}
 \label{detailed_balance_condition_omega}
 \check{\Omega}(\bm{A}'|\bm{A},\bm{Q}) P_{\text{eq}}(\bm{Q},\bm{A})
  = \check{\Omega}(\bm{A}|\bm{A}',\bm{Q}) P_{\text{eq}}(\bm{Q},\bm{A}').
\end{equation}

The dynamics of the coarse-grained system can be fully described by
two hypothetically introduced functions $\check{\Phi}$ and $\check{\Omega}$.
These functions can be interpreted as the trial functions\cite{Schiff-book}.
The optimal forms of these functions should be determined so that they
minimize the differences between the approximate and exact dynamics.
Therefore we can apply the variational method\cite{Schiff-book} to determine
the functional forms of $\check{\Phi}$ and $\check{\Omega}$.
The Kullback-Leibler divergence\cite{Kullback-Leibler-1951} would be
suitable to measure how different two models are\cite{Shell-2008,Espanol-Zuniga-2011}:
\begin{equation}
 \mathcal{K}[\check{\Phi},\check{\Omega}]
  \equiv \int \mathcal{D}\bm{Q} \,
  \check{\mathcal{P}}[\bm{Q} | \check{\Phi},\check{\Omega}]
  \ln \frac{\check{\mathcal{P}}[\bm{Q} | \check{\Phi},\check{\Omega}]}
  {\mathcal{P}[\bm{Q}]} ,
\end{equation}
where $\check{\mathcal{P}}[\bm{Q} | \check{\Phi},\check{\Omega}]$ and
$\mathcal{P}[\bm{Q}]$ are the path probabilities for $\bm{Q}$ by the
approximate and microscopic models.
(The path probability by the approximate dynamics model can be
interpreted as the functional of $\bm{Q}, \check{\Phi}$, and $\check{\Omega}$.)
The Kullback-Leibler divergence
satisfies $\mathcal{K}[\check{\Phi},\check{\Omega}] \ge 0$ and
it becomes zero ($\mathcal{K}[\check{\Phi},\check{\Omega}] = 0$) if two path probabilities are the same.
Therefore, by minimizing the Kullback-Leibler divergence with respect to
trial functions, we have the most reasonable forms for $\check{\Phi}$ and $\check{\Omega}$.
The most reasonable functional forms, $\check{\Phi}^{*}$ and $\check{\Omega}^{*}$,
satisfy the following conditions:
\begin{equation}
 \left. \frac{\delta \mathcal{K}[\check{\Phi},\check{\Omega}]}{\delta \check{\Phi}}\right|_{\check{\Phi}^{*}, \check{\Omega}^{*}} = 0, \qquad
 \left. \frac{\delta \mathcal{K}[\check{\Phi},\check{\Omega}]}{\delta \check{\Omega}}\right|_{\check{\Phi}^{*}, \check{\Omega}^{*}} = 0.
\end{equation}
Unfortunately, the calculation of the path probabilities $\check{\mathcal{P}}[\bm{Q} | \check{\Phi},\check{\Omega}]$ and
$\mathcal{P}[\bm{Q}]$ and the minimization with respect to $\check{\Phi}$ and $\check{\Omega}$
are still not practical. We will need further approximations and simplifications
for the trial functions and the path probabilities. For example, we may assume the
functional form and perform the minimization with respect to several parameters.
We may approximate the path probabilities by the path probability for a single
particle, or we may employ the hypothetical path probability forms based on
dynamical quantities such as the mean-square displacement.

There are several possible simple yet non-trivial models for the dynamics of $\bm{A}$.
Among them, the simplest model would be the following Langevin equation for $\bm{A}$:
\begin{equation}
 \label{langevin_equation_mesoscopic_transient_potential_pseudo}
 \frac{d\bm{A}(t)}{dt} = - \bm{\Gamma} \cdot \frac{\partial \check{\Phi}(\bm{Q}(t),\bm{A}(t))}{\partial \bm{A}(t)}
  + \sqrt{2 k_{B} T} \bm{\Gamma}^{1/2} \cdot \bm{\omega}(t).
\end{equation}
Here, $\bm{\Gamma}$ is the mobility tensor
and $\bm{\omega}$ is the $Z$-dimensional Gaussian white noise.
As before, we have simply assumed that the mobility tensor $\bm{\Gamma}$ is independent of
$\bm{Q}$ and $\bm{A}$.
The fluctuation-dissipation relation
should be satisfied for the noise $\bm{\omega}$:
\begin{equation}
 \langle \bm{\omega}(t) \rangle = 0, \qquad
  \langle \bm{\omega}(t) \bm{\omega}(t') \rangle = \bm{1} \delta(t - t') .
\end{equation}
Eqs~\eqref{langevin_equation_mesoscopic_transient_potential}
and \eqref{langevin_equation_mesoscopic_transient_potential_pseudo}
give the dynamics which is consistent with eqs~\eqref{transition_rate_matrix_approx} and \eqref{detailed_balance_condition_omega}.
This type of coupled Langevin equations correspond to the RaPiD model
for entangled polymers\cite{Kindt-Briels-2007,Briels-2009,Briels-2015}.
We can employ other dynamics models, as well. For example, if the transient potential instantaneously changes,
the simple transition dynamics models would be suitable.
%We assume that the time-dependent joint probability
%distribution of $\bm{A}$ and $\bm{Q}$, $P(\bm{Q},\bm{A};t)$,  obeys the following master equation:
%\begin{equation}
%\begin{split}
% \frac{\partial P(\bm{Q},\bm{A};t)}{\partial t} & =
% \int d\bm{A}' \, \left[ W(\bm{A}|\bm{A}',\bm{Q}) P(\bm{Q},\bm{A}';t)
%  - W(\bm{A}'|\bm{A},\bm{Q}) P(\bm{Q},\bm{A};t) \right] \\
% & \qquad +  \mathcal{L}_{\text{FP}} P(\bm{Q},\bm{A};t),
%\end{split}
%\end{equation}
%where $W(\bm{A}'|\bm{A},\bm{Q})$ is the transition rate from the state $\bm{A}$
%to the state $\bm{A}'$, and $\mathcal{L}_{\text{FP}}$ is the Fokker-Planck operator which
%corresponds to the Langevin equation \eqref{langevin_equation_mesoscopic_transient_potential}:
%\begin{equation}
% \mathcal{L}_{\text{FP}} P(\bm{Q},\bm{A};t)
%  \equiv \frac{\partial}{\partial \bm{Q}^{\mathrm{T}}}
%  \cdot \bm{\Lambda} \cdot
%  \left[ \frac{\partial \Phi(\bm{Q},\bm{A})}{\partial \bm{Q}}
%   P(\bm{Q},\bm{A};t)
%   + k_{B} T \frac{\partial P(\bm{Q},\bm{A};t)}{\partial \bm{Q}}
%   \right] ,
%\end{equation}
%The detailed-balance condition requires the
%transition rate to satisfy
%\begin{equation}
% W(\bm{A}'|\bm{A},\bm{Q}) P_{\text{eq}}(\bm{Q},\bm{A})
%  = W(\bm{A}|\bm{A}',\bm{Q}) P_{\text{eq}}(\bm{Q},\bm{A}') .
%\end{equation}
We may employ a specific transition rate model such as the Glauber dynamics.
Then the transition rate will be explicitly given in terms of the difference
of the transient potential before and after the transition.
This type of coupling of the Langevin equation and transition dynamics corresponds to the
MCSS model\cite{Uneyama-Masubuchi-2012} and the transient bond model\cite{Uneyama-2019} for entangled polymers, and the alternating
diffusive state model for supercooled liquids\cite{Hachiya-Uneyama-Kaneko-Akimoto-2019}.

%------------------------------------------------------------------------------
\section{Discussions}
\label{discussions}

\subsection{Generalized Langevin Equation}
\label{generalized_langevin_equation}

We have proposed the LETP model by introducing the transient potential to
approximately describe the mesoscopic dynamics. Also, we have proposed some
possible approximate dynamics model for the transient potential by introducing the
pseudo thermodynamic degrees of freedom.
This is not a unique way to describe
the complex mesoscopic dynamics. We may employ other methods to describe
the mesoscopic dynamics. The most popular and established way is to use
the projection operator\cite{Kawasaki-1973,Dengler-2016}. The projection operator method gives the 
GLE as the effective dynamic equation for the mesoscopic degrees of freedom.
The GLE involves the memory kernel which directly
expresses the memory effect for the mesoscopic degrees of freedom.
In this subsection, we compare the LETP model with the dynamic equation
which incorporates the memory kernel.

We start from the same microscopic dynamics model as Sec.~\ref{theory}, and
consider the effective dynamic equation for the degrees of freedom $\bm{Q}$.
By eliminating the fast degrees of freedom, we have the
GLE as the dynamic equation
for the mesoscopic degrees of freedom:
\begin{equation}
 \label{generalized_langevin_equation_mesoscopic}
 \frac{d\bm{Q}(t)}{dt}
  = - \int_{-\infty}^{t} dt' \, \bm{K}(t - t') \cdot \frac{\partial \mathcal{F}(\bm{Q}(t'))}{\partial \bm{Q}(t')}
  + \bm{\xi}(t) ,
\end{equation}
where $\bm{K}(t)$ is the memory kernel and $\bm{\xi}(t)$ is the colored noise.
The fluctuation-dissipation relation requires the noise to satisfy
\begin{equation}
 \label{fluctuation_dissipation_relation_mesoscopic}
 \langle \bm{\xi}(t) \rangle = 0, \qquad
  \langle \bm{\xi}(t) \bm{\xi}(t') \rangle = k_{B} T \bm{K}(|t - t'|).
\end{equation}
The projection operator method gives eqs~\eqref{generalized_langevin_equation_mesoscopic}
and \eqref{fluctuation_dissipation_relation_mesoscopic}, but it does not tell
us the detailed statistical properties of the colored noise $\bm{\xi}(t)$.
In most practical cases, the colored noise $\bm{\xi}(t)$ is simply assumed to be
Gaussian. (This assumption seems to be often employed implicitly.) Then the dynamic
equation for the mesoscopic degrees of freedom can be fully specified.
This Gaussian assumption cannot be justified {\it a priori}, and we should interpret it as an approximation.
In this work, we explicitly distinguish the GLE with the Gaussian noise (GLEG) with
the GLE with a general non-Gaussian noise.
It would be reasonable to consider that both the GLEG and the LETP can be obtained
from the same microscopic dynamics model with different approximations. We expect that
the difference between the GLEG and the LETP originates from the properties
of the employed approximations.

To consider the difference between the GLEG and the LETP in detail, it
would be better for us to derive the GLEG by utilizing the path probability
and the Onsager-Machlup action. Therefore here we go back to eqs~\eqref{path_probability_position_mesoscopic}
and \eqref{onsager_machlup_action_mesoscopic_modified}.
As we mentioned, two actions in eq~\eqref{onsager_machlup_action_mesoscopic_modified}
are coupled via the interaction potential $U(\bm{Q},\bm{\theta})$.
In the derivation of the LETP, we introduced the transient potential to
rewrite the action for $\bm{Q}$ in a simple form.
Here we consider to introduce a different quantity to simplify the action for $\bm{Q}$.
We consider an average of the force term for $\bm{Q}$,
\begin{equation}
 \label{average_velocity_mesoscopic}
 \bar{\bm{v}}[\bm{Q},t] = 
 - \left\langle \bm{\Lambda} \cdot \frac{\partial U(\bm{Q}(t),\bm{\theta}(t))}{\partial \bm{Q}(t)}
   \right\rangle^{(\bm{\theta})} ,
\end{equation}
where $\langle \dots \rangle^{(\bm{\theta})}$ represents the statistical average over $\bm{\theta}$.
The thus defined $\bar{\bm{v}}$ can be interpreted as the average ``velocity'' for the
mesoscopic degrees of freedom $\bm{Q}$. From the causality, $\bar{\bm{v}}$ at time $t$ is a functional of $\bm{Q}(t')$ for $t' \le t$.
If the system is fluctuating around the equilibrium, $\bar{\bm{v}}$ should be expressed as
a linear function of the thermodynamic force. Thus we expect the following form for $\bar{\bm{v}}$:
\begin{equation}
 \label{average_velocity_mesoscopic_by_thermodynamic_force}
 \bar{\bm{v}}[\bm{Q},t] = - \int_{-\infty}^{t} dt' \,
  \bm{K}(t - t') \cdot \frac{\partial \mathcal{F}(\bm{Q}(t'))}{\partial \bm{Q}(t')} .
\end{equation}
We may employ eq~\eqref{average_velocity_mesoscopic_by_thermodynamic_force} as
the definition of $\bar{\bm{v}}$, instead of
eq~\eqref{average_velocity_mesoscopic}.
Anyway, $\bar{\bm{v}}$ is an average and the force term is fluctuating around it.
We introduce the deviation of the force term from
$\bar{\bm{v}}$ as $\Delta \bm{v}(t)$:
\begin{equation}
 \Delta \bm{v}(t) =
  - \bm{\Lambda} \cdot \frac{\partial U(\bm{Q}(t),\bm{\theta}(t))}{\partial \bm{Q}(t)}
  - \bar{\bm{v}}[\bm{Q},t] .
\end{equation}
This $\Delta \bm{v}(t)$ can be interpreted as the fluctuation around the reference path.
As before, we utilize the functional identity to introduce $\Delta \bm{v}$ as
additional degrees of freedom:
\begin{equation}
 \label{delta_functional_identity_velocity}
 1 = \int \mathcal{D}\Delta \bm{v} \,
  \delta\left[  \Delta \bm{v}(t) -
 \bm{\Lambda} \cdot \frac{\partial U(\bm{Q}(t),\bm{\theta}(t))}{\partial \bm{Q}(t)}
  - \bar{\bm{v}}[\bm{Q},t]
 \right] .
\end{equation}
We insert eq~\eqref{delta_functional_identity_velocity}
into eq~\eqref{path_probability_position_mesoscopic}. Then we can rewrite
the path probability for $\bm{Q}$ as
\begin{equation}
 \label{effective_path_probability_position_mesoscopic_memory}
  \begin{split}
   \mathcal{P}[\bm{Q}]
   & = \int \mathcal{D}\bm{\theta} \mathcal{D}\Delta \bm{v} \,
  \delta\left[  \Delta \bm{v}(t) -
 \bm{\Lambda} \cdot \frac{\partial U(\bm{Q}(t),\bm{\theta}(t))}{\partial \bm{Q}(t)}
  - \bar{\bm{v}}[\bm{Q},t]
 \right] \mathcal{N}^{(\bm{Q},\bm{\theta})}
  \exp\left[ - \mathcal{S}[\bm{Q},\bm{\theta}] \right] \\
   & = \int \mathcal{D}\Delta\bm{v} \, \mathcal{N}^{(\bm{Q},\Delta \bm{v})}
  \exp\left[ - \bar{\mathcal{S}}^{(\bm{Q})}[\bm{Q}|\Delta \bm{v}]
   - \bar{\mathcal{S}}^{(\Delta \bm{v})}[\Delta \bm{v}|\bm{Q}]   \right] ,
  \end{split}
\end{equation}
with
\begin{equation}
 \label{onsager_machlup_action_mesoscopic_memory}
   \bar{\mathcal{S}}^{(\bm{Q})}[\bm{Q}|\Delta \bm{v}]
    \equiv \frac{1}{2 k_{B} T} \int dt \,
  G\left(\frac{d \bm{Q}(t)}{dt} - \bar{\bm{v}}[\bm{Q},t] - \Delta \bm{v}(t)
   ;  \bm{\Lambda}
      \right) ,
\end{equation}
\begin{equation}
 \label{onsager_machlup_action_velocity_fluctuation_memory}
  \begin{split}
   \bar{\mathcal{S}}^{(\Delta \bm{v})}[\Delta \bm{v}|\bm{Q}]
   & \equiv - \ln \int \mathcal{D}\bm{\theta} \,   \delta\left[  \Delta \bm{v}(t) -
 \bm{\Lambda} \cdot \frac{\partial U(\bm{Q}(t),\bm{\theta}(t))}{\partial \bm{Q}(t)}
  - \bar{\bm{v}}[\bm{Q},t]
 \right] \\
   & \qquad \times \exp\left[ - \frac{1}{2 k_{B} T} \int dt \,
   G\left(\frac{d \bm{\theta}(t)}{dt} + \bm{M} \cdot \frac{\partial U(\bm{Q}(t),\bm{\theta}(t))}{\partial \bm{\theta}(t)}
      ; \bm{M} \right) 
  \right] ,
  \end{split}
\end{equation}
where $\mathcal{N}^{(\bm{Q},\Delta\bm{v})}$ is the normalization factor.
Eqs~\eqref{effective_path_probability_position_mesoscopic_memory}-\eqref{onsager_machlup_action_velocity_fluctuation_memory}
have similar forms to eqs~\eqref{effective_path_probability_position_mesoscopic_modified}-\eqref{onsager_machlup_action_remaining}.
As the case of eqs~\eqref{effective_path_probability_position_mesoscopic_modified}-\eqref{onsager_machlup_action_remaining},
eqs~\eqref{effective_path_probability_position_mesoscopic_memory}-\eqref{onsager_machlup_action_velocity_fluctuation_memory}
are derived without approximations and thus they are formally exact.

To obtain the GLEG, we approximate the action for $\Delta \bm{v}$
(eq~\eqref{onsager_machlup_action_velocity_fluctuation_memory}) by a simple Gaussian form:
\begin{equation}
 \label{onsager_machlup_action_velocity_fluctuation_memory_approx}
   \bar{\mathcal{S}}^{(\Delta \bm{v})}[\Delta \bm{v}|\bm{Q}]
    \approx  - \frac{1}{2 k_{B} T} \int dt dt' \,
   \Delta \bm{v}^{\mathrm{T}}(t) \cdot \bar{\bm{C}}^{-1}(t - t') \cdot \Delta \bm{v}(t') ,
\end{equation}
where $\bar{\bm{C}}(t)$ is a tensor which represents the covariance of $\Delta \bm{v}$.
(The explicit form of this tensor is not required here.)
Under this approximation, the path probability for $\bm{Q}$ can be explicitly calculated.
The Gaussian weight for $\bm{Q}$ in eq~\eqref{onsager_machlup_action_mesoscopic_memory}
can be also interpreted as a Gaussian weight for $\Delta \bm{v}$.
Thus the path probability for $\bm{Q}$ can be calculated by integrating the
path probability over $\Delta\bm{v}$. 
From eqs~\eqref{effective_path_probability_position_mesoscopic_memory},
~\eqref{onsager_machlup_action_mesoscopic_memory}, and 
\eqref{onsager_machlup_action_velocity_fluctuation_memory_approx}, we have
\begin{equation}
 \label{effective_path_probability_position_mesoscopic_memory_approx}
  \begin{split}
   \mathcal{P}[\bm{Q}]
   & \approx \int \mathcal{D}\Delta\bm{v} \, \mathcal{N}^{(\bm{Q},\Delta \bm{v})}
  \exp\bigg[  - \frac{1}{2 k_{B} T} \int dt dt' \, \bigg[
   \Delta \bm{v}^{\mathrm{T}}(t) \cdot \bar{\bm{C}}^{-1}(t - t') \cdot \Delta \bm{v}(t')
 \\
   & \qquad +
  \left(\Delta \bm{v}(t) - \frac{d \bm{Q}(t)}{dt} + \bm{v}[\bm{Q},t] 
      \right)^{\mathrm{T}}
   \cdot 2 \bm{\Lambda} \delta(t - t') \cdot
  \left(\Delta \bm{v}(t') - \frac{d \bm{Q}(t')}{dt'} + \bm{v}[\bm{Q},t'] 
      \right)
   \bigg]  \bigg] \\
   & = \mathcal{N}^{(\bm{Q})}
  \exp\bigg[  - \frac{1}{2 k_{B} T} \int dt dt' \, 
  \left(\frac{d \bm{Q}(t)}{dt} - \bm{v}[\bm{Q},t] 
      \right)^{\mathrm{T}}\cdot \bar{\bm{K}}^{-1}(t - t') \cdot
  \left(\frac{d \bm{Q}(t')}{dt'} - \bm{v}[\bm{Q},t'] 
      \right)
   \bigg]  ,
  \end{split}
\end{equation}
where $ \mathcal{N}^{(\bm{Q})}$ is the normalization factor and $\bar{\bm{K}}(t)$ is the kernel function
defined as
\begin{equation}
 \bar{\bm{K}}(t) = \bar{\bm{C}}(t) + 2 \bm{\Lambda} \delta(t) .
\end{equation}
Eq~\eqref{effective_path_probability_position_mesoscopic_memory_approx}
is equivalent to the GLEG if the kernel $\bar{\bm{K}}(t)$ is given as
$\bar{\bm{K}}(t) = \bm{K}(|t|)$. This condition is equivalent to
the fluctuation-dissipation relation \eqref{fluctuation_dissipation_relation_mesoscopic},
 and thus it should be satisfied to reproduce the
correct equilibrium distribution.
Thus we find that the GLEG can be obtained from 
eqs~\eqref{path_probability_position_mesoscopic}
and \eqref{onsager_machlup_action_mesoscopic_modified}, if we approximate
the fluctuation of the force term $\Delta\bm{v}$ by a simple Gaussian
from (eq~\eqref{onsager_machlup_action_velocity_fluctuation_memory_approx}).

By comparing the derivations of the GLEG and the LETP, we find some
differences between them. The first difference is that the LETP employs
additional degrees of freedom, the transient potential, to express the
force term in the action \eqref{onsager_machlup_action_mesoscopic_modified}.
The GLEG employs the average $\bar{\bm{v}}$, which is a functional of $\bm{Q}$,
instead. This average $\bar{\bm{v}}$ incorporates the memory kernel.
The second difference is that the additional degrees of freedom is not eliminated
in the LETP. In other words, we explicitly have the dynamic equation
for the additional degrees of freedom (the transient potential $\Phi$), in addition to that for the mesoscopic
degrees of freedom $\bm{Q}$.
This is in contrast to the case of the GLEG. To derive the GLEG, we eliminated
the fluctuation around the average, $\Delta \bm{v}$, by integrating the 
path probability over it.
The LETP does not require the memory kernel but requires additional degrees of freedom,
whereas the GLEG does not require additional degrees of freedom but requires
the memory kernel.

\subsection{Example: Supercooled Liquid}
\label{example_supercooled_liquid}

Because the GLEG and the LETP are based on different approximations,
some statistical properties of them can be quantitatively different, although the
target system is the same.
As a simple example, here we consider the effective dynamic equation model for a single
tagged particle (or the center of mass of a tagged molecule) in a supercooled
liquid in a three dimensional space.

The dynamics of supercooled liquids have been widely studied by binary Lennard-Jones mixture
systems\cite{Kob-Andersen-1994,Kob-Andersen-1995,Kob-Andersen-1995a,Yamamoto-Onuki-1998,Yamamoto-Onuki-1998a,Kob-1999,Vorselaars-Lyulin-Karatasos-Michels-2007}.
To study the diffusion behavior, the mean-square displacement (MSD) data are useful.
If the temperature is sufficiently high, we just observe normal diffusion
behavior: $\langle [\bm{r}(t) - \bm{r}(0)]^{2} \rangle \propto t$.
If the temperature is sufficiently low, we observe the slowing-down of the
dynamics. As a result,
the MSD of the particle typically show three regions\cite{Hunter-Weeks-2012,Klix-Maret-Keim-2015}.
At the short time region, it exhibits a normal diffusion. At the intermediate
time region, the MSD becomes almost independent of time, and exhibits a
plateau. At the long time region, it again exhibits a normal diffusion.
Therefore, the MSD of a particle would be as follows:
\begin{equation}
 \label{msd_supercooled_liquid_scaling}
 \langle [\bm{r}(t) - \bm{r}(0)]^{2} \rangle
  \propto
  \begin{cases}
   t^{1} & (t \lesssim \tau') , \\
   t^{0} & (\tau' \lesssim t \lesssim \tau'') , \\
   t^{1} & (\tau'' \lesssim t) ,
  \end{cases}
\end{equation}
where $\tau'$ and $\tau''$ are characteristic time scales.
(Strictly speaking, at sufficiently short time scale, we observe the ballistic
diffusion behavior. In this work we consider overdamped dynamics and thus
we do not consider the ballistic region.)
The MSD data are not sufficient to characterize the dynamics of a particle.
The distribution of the displacement is generally not Gaussian, and the non-Gaussianity
cannot be detected via the MSD.
The non-Gaussianity parameter (NGP)\cite{Rahman-1964,Vorselaars-Lyulin-Karatasos-Michels-2007}, which characterizes the deviation
of the diffusion behavior from the ideal Gaussian behavior, is useful to
study the non-Gaussianity. For a three dimensional system, the NGP is defined
as $\alpha(t) \equiv 3 \langle |\bm{r}(t) - \bm{r}(0)|^{4} \rangle / 5 \langle [\bm{r}(t) - \bm{r}(0)]^{2} \rangle^{2} - 1 $.

\begin{figure}[h]
\includegraphics[width=.8\linewidth,clip]{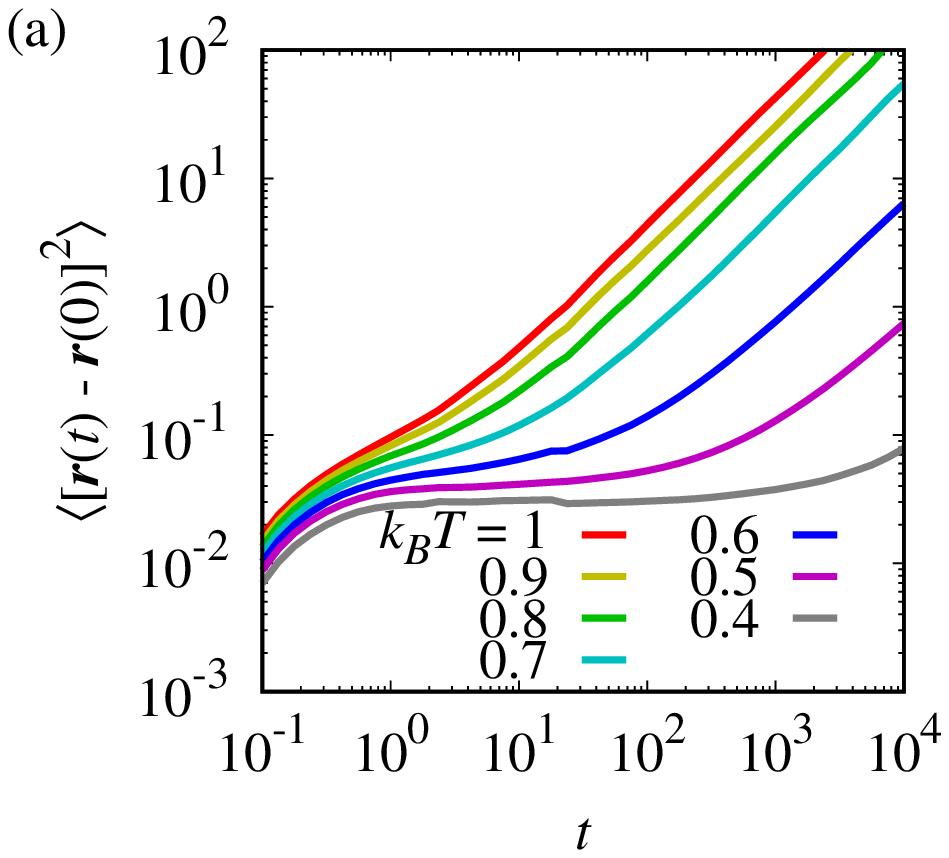}
\includegraphics[width=.8\linewidth,clip]{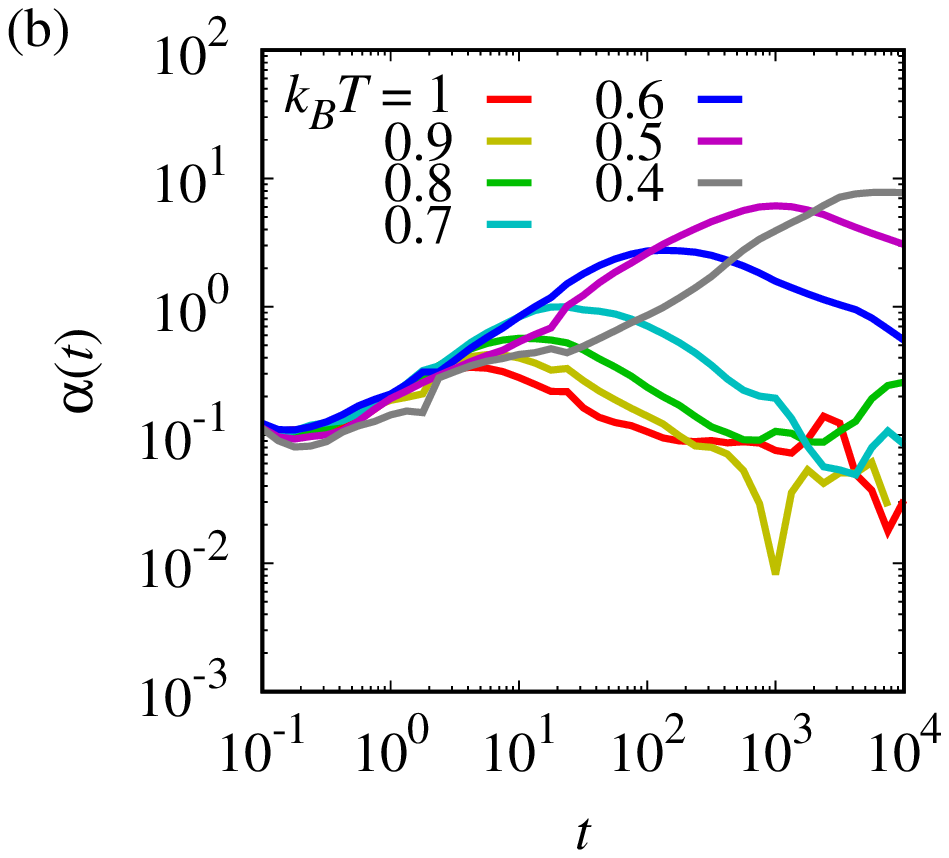}
\caption{\label{md_msd_and_ngp}}
\end{figure}

\begin{figure}[h]
\includegraphics[width=.4\linewidth,clip]{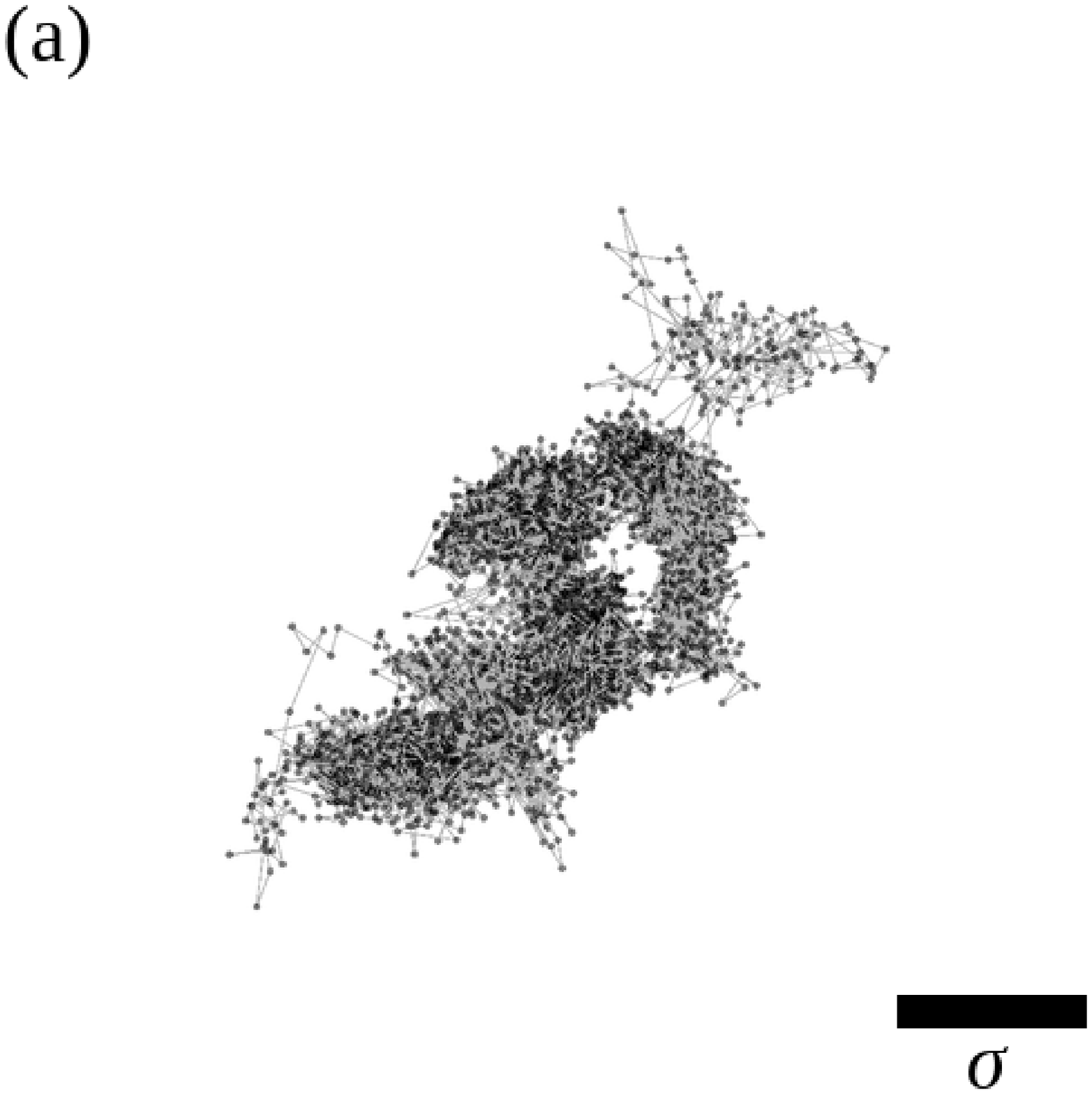}%
\includegraphics[width=.4\linewidth,clip]{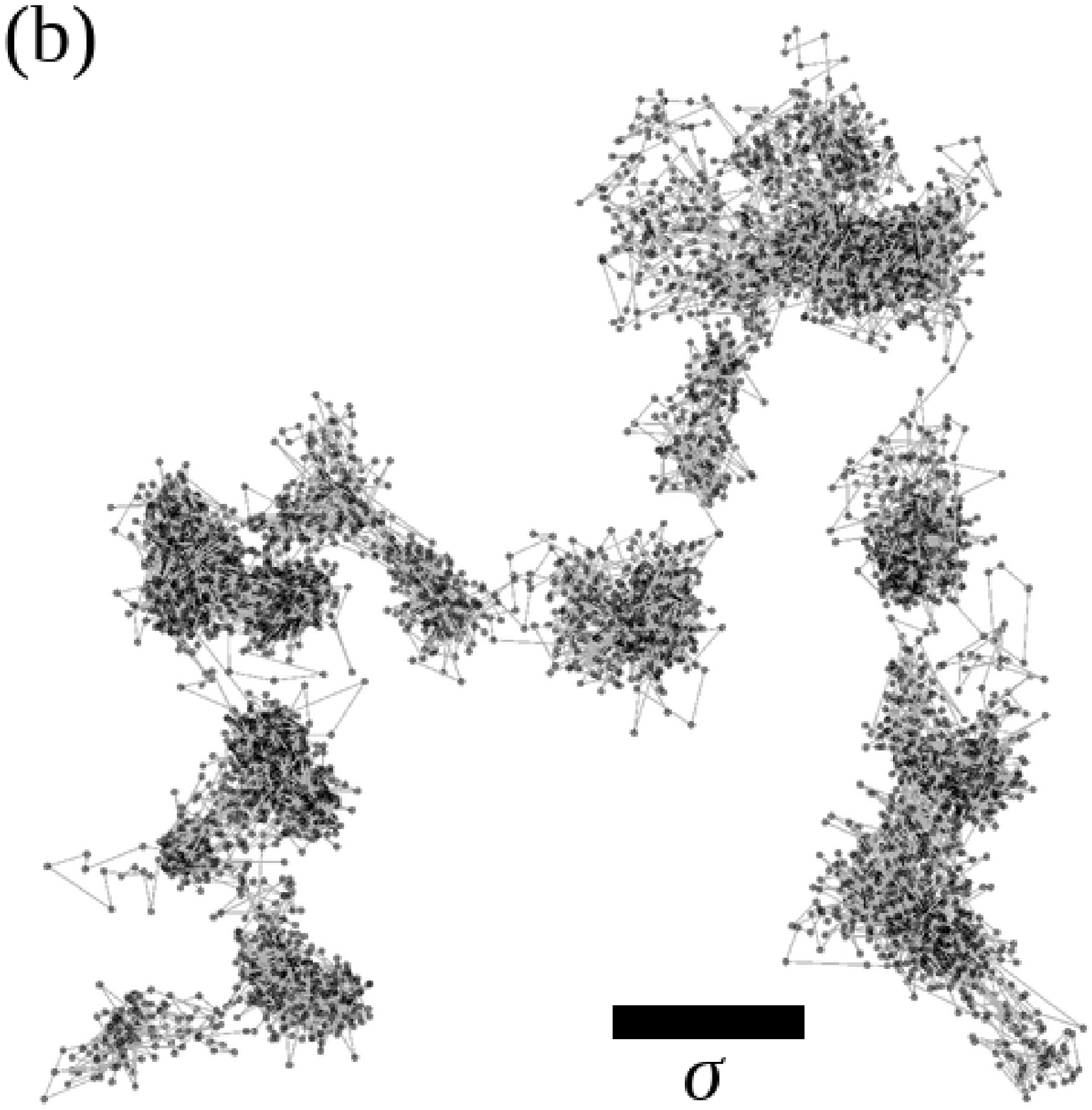}
\includegraphics[width=.4\linewidth,clip]{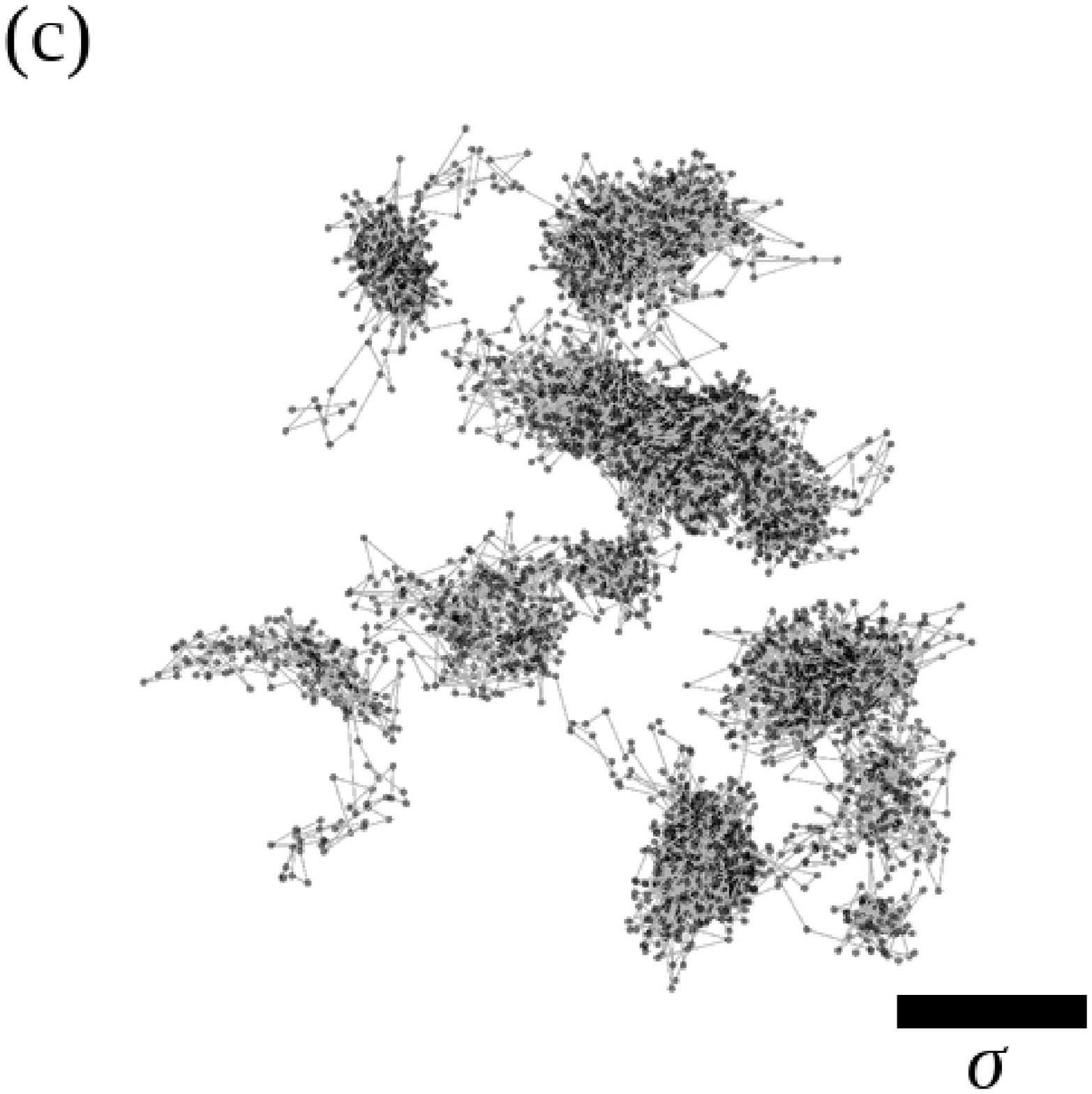}%
\includegraphics[width=.4\linewidth,clip]{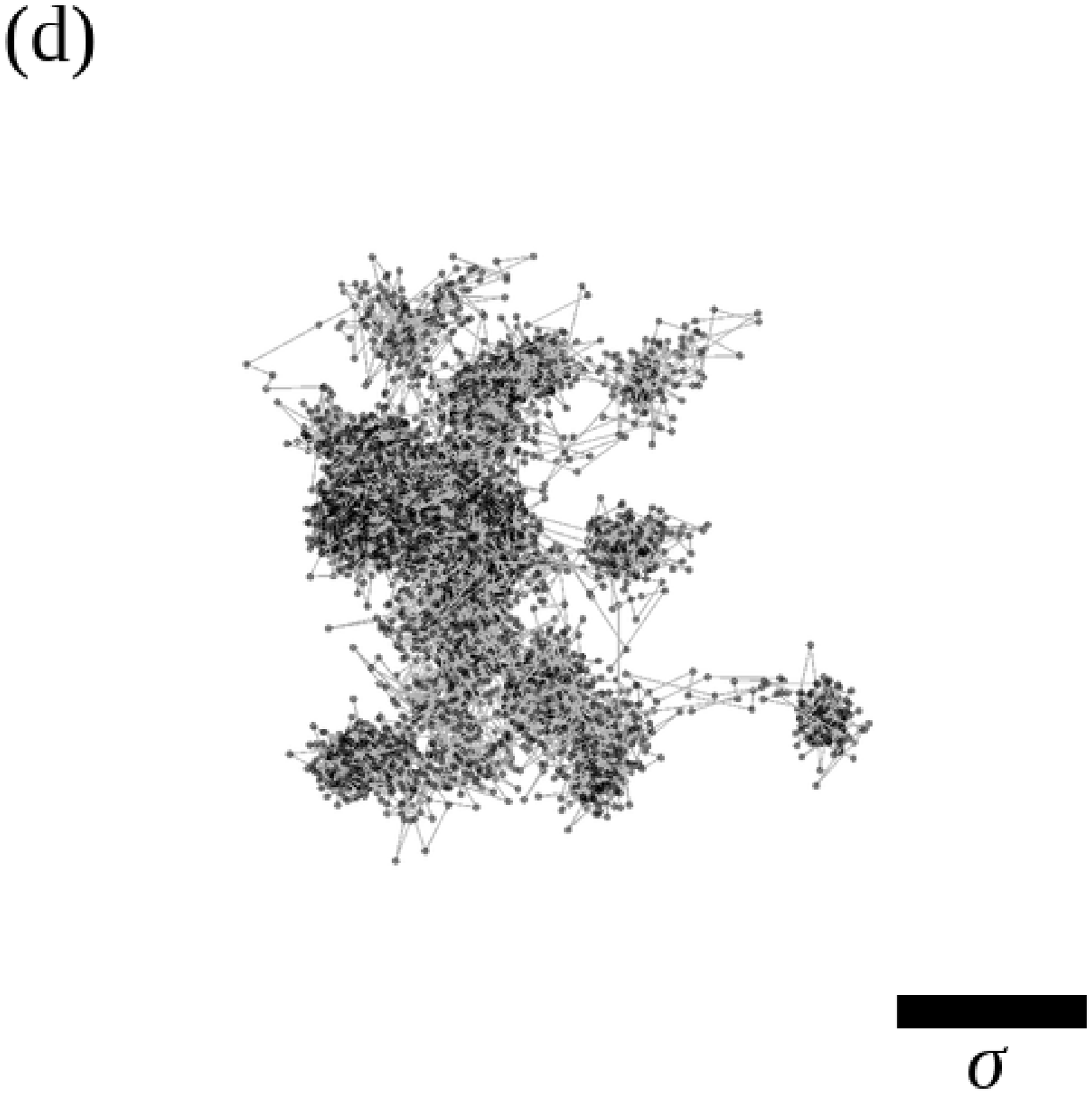}
\caption{\label{md_trajectories}}
\end{figure}

We show the MSD and NGP of a particle in a model binary Lennard-Jones mixture with different temperatures in Figure~\ref{md_msd_and_ngp}.
Here, the dimensionless units are employed (the characteristic length, mass, and energy are set to be unity) and
the temperature is changed from $k_{B} T = 0.4$ to $k_{B} T = 1$.
The details of the simulation model and the simulation setup are shown in Appendix~\ref{molecular_dynamics_simulation_for_supercooled_liquid}.
We show some trajectories of particles in a supercooled
liquid at $k_{B} T = 0.6$ in Figure~\ref{md_trajectories}. We clearly observe that the trajectories are qualitatively
different from those of normal Brownian motions. This can be interpreted as
the fluctuation of the mobility, which is called the dynamic heterogeneity.

We consider whether such behavior can be successfully modeled
by the GLEG and the LETP.
We express the position of the tagged particle as $\bm{r}(t)$, and use
this as the mesoscopic degrees of freedom.
We construct the effective dynamic equations for $\bm{r}$, and then
analyze the MSD and NGP.

From the translational symmetry,
the free energy is zero: $\mathcal{F}(\bm{r}) = 0$.
Therefore, if we employ the GLEG to describe the dynamics, we have the dynamic equation
as
\begin{equation}
 \label{gleg_supercooled_liquid}
 \frac{d\bm{r}(t)}{dt} = \bm{\xi}(t) ,
\end{equation}
where $\bm{\xi}(t)$ is the Gaussian colored noise. The first and second
moments of the noise $\bm{\xi}(t)$ are
\begin{equation}
 \langle \bm{\xi}(t) \rangle = 0, \qquad
 \langle \bm{\xi}(t) \bm{\xi}(t') \rangle = k_{B} T K(|t - t'|) \bm{1}.
\end{equation}
Here $K(t)$ is the (scalar) memory kernel. We have assumed that the system is
isotropic and the memory kernel tensor is given as an isotropic tensor.
The MSD is simply calculated to be
\begin{equation}
 \label{gleg_msd_supercooled_liquid}
 \langle [\bm{r}(t) - \bm{r}(0)]^{2} \rangle
  = \int_{0}^{t} dt' \int_{0}^{t} dt'' \, \langle \bm{\xi}(t') \cdot \bm{\xi}(t'') \rangle
  = 6 k_{B} T \int_{0}^{t} dt' \, (t - t') K(t').
\end{equation}
Thus we find that the memory kernel can be determined if the MSD of the
tagged particle is given.
From the Gaussian nature of
the noise $\bm{\xi}(t)$, the NGP is exactly zero: $\alpha(t) = 0$.
%\begin{equation}
% \alpha(t) \equiv \frac{3  \langle |\bm{r}(t) - \bm{r}(0)|^{4} \rangle}
%  {5 \langle [\bm{r}(t) - \bm{r}(0)]^{2} \rangle^{2}} - 1 = 0.
%\end{equation}
This means that, the GLEG can reproduce the MSD observed in supercooled
liquids successfully (by tuning the memory kernel), but it cannot reproduce the non-Gaussian behavior.

\begin{figure}[h]
\includegraphics[width=.8\linewidth,clip]{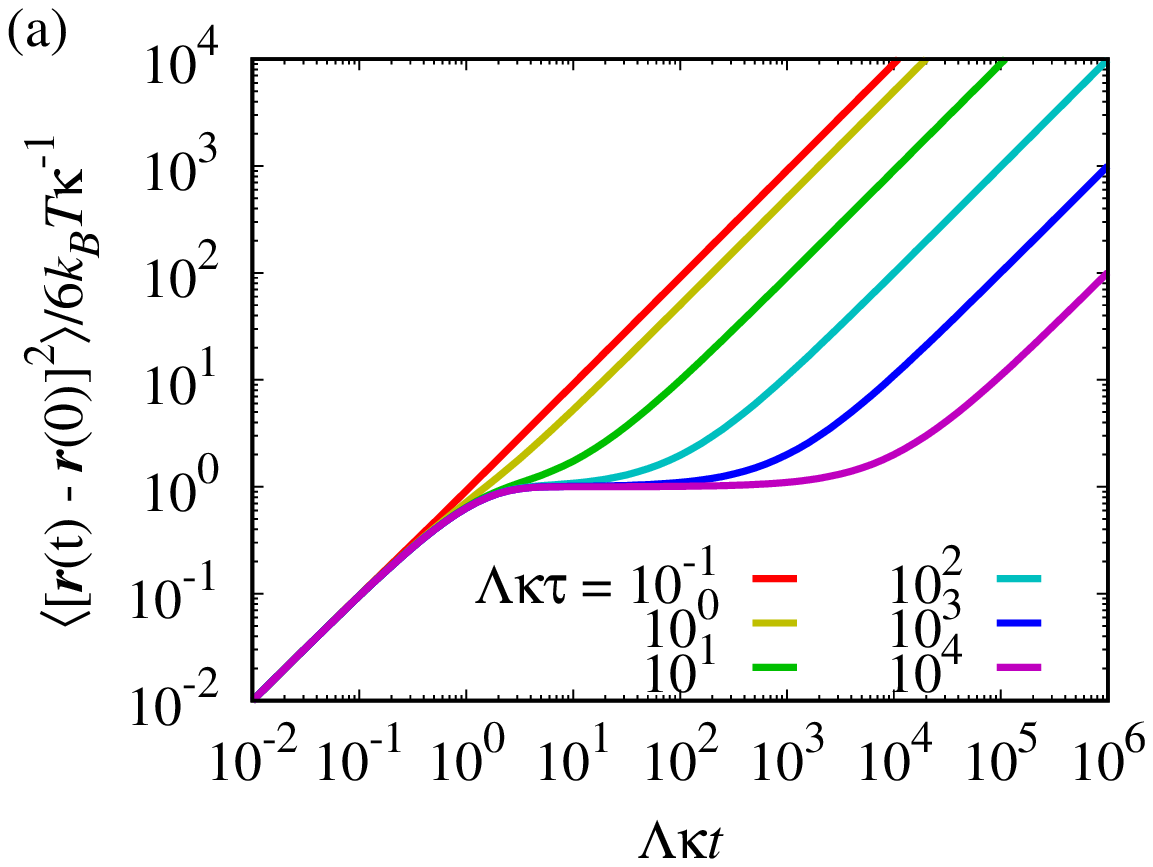}
\includegraphics[width=.8\linewidth,clip]{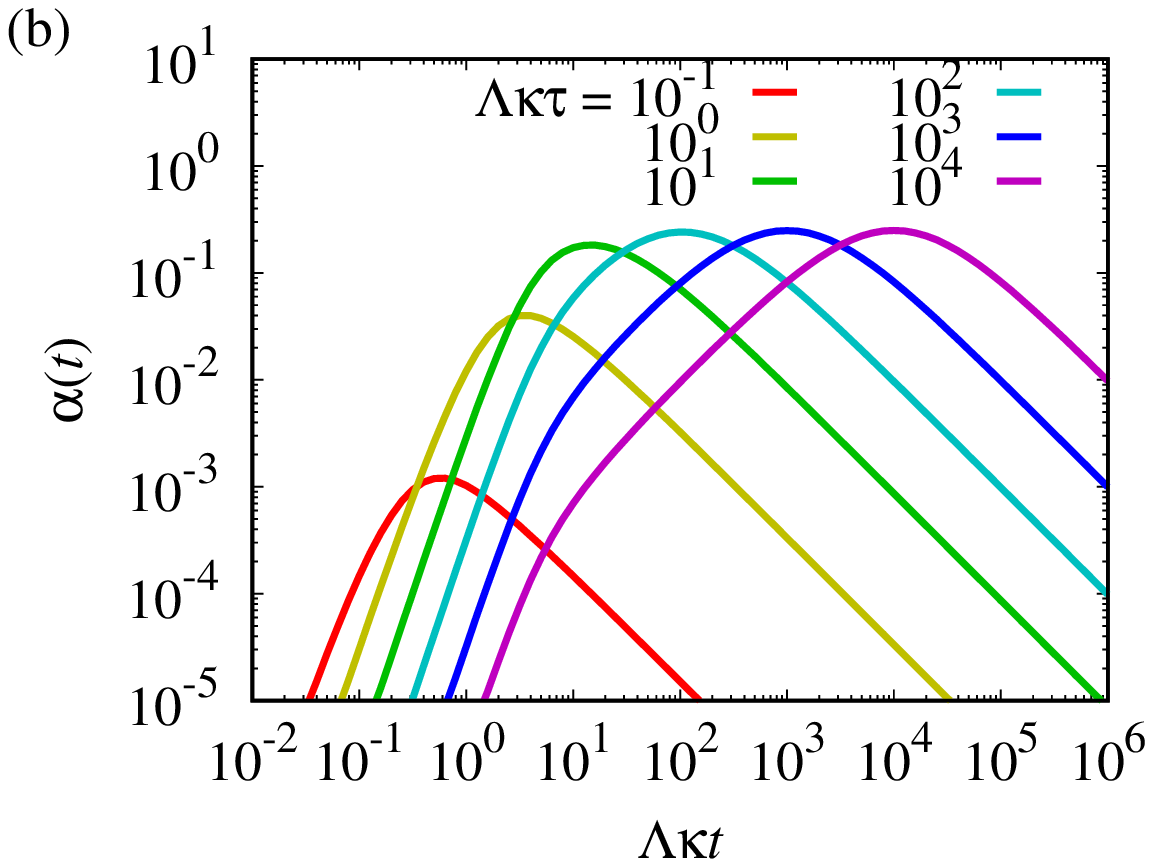}
\caption{\label{letp_msd_and_ngp}}
\end{figure}

If we employ the LETP, the dynamic equation becomes
\begin{equation}
 \label{langevin_equation_transient_potential_supercooled_general}
 \frac{d\bm{r}(t)}{dt} = - \Lambda \frac{\partial \check{\Phi}(\bm{r}(t),\bm{A}(t))}{\partial \bm{r}(t)}
  + \sqrt{2 k_{B} T \Lambda} \bm{W}(t) ,
\end{equation}
where $\Lambda$ is the (scalar) mobility and $\bm{W}(t)$ is the Gaussian white
noise. The force term by the transient potential in eq~\eqref{langevin_equation_transient_potential_supercooled_general} is
not zero.
Unlike the case of the GLEG, we should specify the dynamics model
of the transient potential or the pseudo thermodynamic degrees of freedom.
As a simple yet nontrivial model, we employ a simple harmonic type potential as the transient potential:
\begin{equation}
 \label{transient_potential_supercooled}
  \check{\Phi}(\bm{r},\bm{A}) = \frac{1}{2} \kappa (\bm{r} - \bm{A})^{2} ,
\end{equation}
where $\kappa$ is the spring constant and $\bm{A}$ corresponds to the center position of the potential.
(The dimension of $\bm{A}$ is assumed to be
the same as that of $\bm{r}$.) The dynamic equation
can be then simplified as
\begin{equation}
 \label{langevin_equation_transient_potential_supercooled}
 \frac{d\bm{r}(t)}{dt} = - \Lambda \kappa [\bm{r}(t) - \bm{A}(t)]
  + \sqrt{2 k_{B} T \Lambda} \bm{W}(t) .
\end{equation}
We need to specify the dynamics model for the pseudo thermodynamic degrees of freedom.
If we employ the Langevin equation for $\bm{A}(t)$, the full stochastic process
become a Gaussian process, and thus the results will be very similar to those
of the GLEG. Namely, the MSD will be reproduced but the NGP is always zero.
Here we employ the stochastic transition dynamics with the following transition rate, instead:
\begin{equation}
 \label{transition_rate_transient_potential_supercooled}
  \check{\Omega}(\bm{A}'|\bm{A},\bm{r})
  = \frac{1}{\tau} \left(\frac{\kappa}{2 \pi k_{B} T}\right)^{3/2}
  \exp 
  \left[ - \frac{\kappa (\bm{A}' - \bm{r})^{2}}{2 k_{B} T} \right] ,
\end{equation}
where $\tau$ is the characteristic time of the transition. This transition
rate model corresponds to the simple resampling of the new potential center
position from the equilibrium probability distribution.
Now the dynamics of the system can be fully specified by
eqs~\eqref{langevin_equation_transient_potential_supercooled}
and \eqref{transition_rate_transient_potential_supercooled}.
(This model would be interpreted as a special case of the alternating
diffusive state model\cite{Hachiya-Uneyama-Kaneko-Akimoto-2019}, where the fraction of the free diffusive state is very small.)
Although the model looks simple, the calculations of the MSD and NGP become
rather complicated. We show the detailed calculations in Appendix~\ref{detailed_calculation_for_msd_and_ngp}, and
here we only show the results. The MSD and NGP of our model become
\begin{equation}
 \label{msd_expression_supercooled_final}
 \langle [\bm{r}(t) - \bm{r}(0)]^{2} \rangle
  = \frac{6 k_{B} T}{\kappa} \frac{  \eta }{1 + \eta}
  \left[  \frac{t}{\tau}
  + \frac{\eta}{1 + \eta} 
  [ 1 - e^{- t (1 + \eta) / \tau}] \right] ,
\end{equation}
\begin{equation}
 \label{ngp_expression_supercooled_final}
  \begin{split}
   \alpha(t)
   & = \left[ \frac{t}{\tau}
  + \frac{\eta  [ 1 - e^{- t (1 + \eta) / \tau}]}{1 + \eta}  \right]^{-2}     \bigg[ 
  \frac{2 \eta^{2}}{(1 + \eta) (1 + 2 \eta)} 
  \frac{t}{\tau}
  + \frac{4 \eta}{1 + \eta} 
  \frac{t}{\tau} e^{- t (1 + \eta) / \tau}
  \\
  & \qquad
   + \frac{4 [ 1  - e^{-t (1 + \eta) / \tau} ]}{(1 + \eta)^{2}}
   - \frac{4 (1 +\eta)^{2}   [ 1 - e^{-t (1 + 2 \eta) / \tau} ]}{(1 + 2 \eta)^{2} } 
   + \frac{\eta^{2} [ 1 - e^{- 2 t (1 + \eta) / \tau}]}{(1 + \eta)^{2}} 
   \bigg] ,
  \end{split}
\end{equation}
where $\eta \equiv \Lambda \kappa \tau$.
We show the MSD and NGP data by the LETP, with various average waiting times,
in Figure~\ref{letp_msd_and_ngp}. If the waiting time is sufficiently short,
the transient potential does not contribute the diffusion dynamics.
Thus, in the case of $\Lambda \kappa \tau \ll 1$, we recover the simple
diffusion behavior where the MSD is proportional to $t$ and the NGP is almost zero.
On the other hand, if the waiting time is sufficiently long, the particle
will be trapped in the transient potential and exhibits the plateau at the
intermediate region. The MSD data by the LETP are qualitatively consistent
with the data by the molecular dynamics simulation, Figure~\ref{md_msd_and_ngp}(a).
For example, eq~\eqref{msd_expression_supercooled_final} clearly exhibits three regions
shown in eq~\eqref{msd_supercooled_liquid_scaling}.
In addition, the LETP gives non-zero NGP. Although the $t$-dependence of the NGP
by the LETP is not quantitatively coincide with that by the molecular dynamics simulation,
the trend is qualitatively reproduced by the LETP.
In both Figures~\ref{md_msd_and_ngp}(b) and
\ref{letp_msd_and_ngp}(b), the NGP exhibits a peak
where the MSD shows the crossover from the plateau to the diffusion behavior.
The peak value of the NGP increases as the plateau region
in the MSD develops.

By comparing the results of the GLEG and the LETP, we find that
the MSD can be well described both
by the GLEG and the LETP.
The GLEG can easily reproduce any MSD by tuning the
memory kernel. However, the diffusion dynamics given by the GLEG
is essentially a Gaussian process and non-Gaussian behavior can never be reproduced.
On the other hand, the LETP can reasonably reproduce the non-Gaussian behavior.
But both the MSD and NGP depend on the dynamics model and
the tuning of the forms of a transient potential and
a dynamics model such as a transition rate is difficult.

The simple structure of the LETP would be especially useful when we perform
numerical simulations. According to the results shown above, the LETP
model can successfully reproduce some dynamical properties of supercooled liquids.
If we integrate such a dynamics model into more complex systems,
we will be able to simulate complex relaxation process with a relatively
simple and numerically efficient model. For example, if we combine the
single chain polymer model (such as the Rouse model) with the LETP in this subsection, we may be
able to simulate the dynamics of supercooled polymer melts by a simple
single chain model.

\subsection{Fluctuating Diffusivity}
\label{fluctuating_diffusivity}

In Secs.~\ref{generalized_langevin_equation} and \ref{example_supercooled_liquid},
we have showed that the LETP is qualitatively different from the GLEG.
Recently, another type of mesoscopic coarse-grained model which is
called the fluctuating diffusivity (or diffusing diffusivity) model has
been investigated. In this model, the diffusion coefficient tensor (or the
mobility tensor) is considered as a stochastically fluctuating physical
quantity. The dynamic equation is expressed as the Langevin equation
with the fluctuating diffusivity (LEFD)\cite{Uneyama-Miyaguchi-Akimoto-2015,Miyaguchi-Akimoto-Yamamoto-2016,Miyaguchi-2017,Uneyama-Miyaguchi-Akimoto-2019,Miyaguchi-Uneyama-Akimoto-2019}. The LEFD for the mesoscopic
degrees of freedom $\bm{Q}$ can be expressed as
\begin{equation}
 \label{langevin_equation_mesoscopic_fluctuating_diffusivity}
 \frac{d\bm{Q}(t)}{dt}
  = - \frac{1}{k_{B} T} \bm{D}(t) \cdot \frac{\partial \mathcal{F}(\bm{Q}(t))}{\partial \bm{Q}(t)}
  + \sqrt{2} \bm{D}^{1/2}(t) \cdot \bm{W}(t),
\end{equation}
where $\bm{D}(t)$ is the time-dependent fluctuating diffusion coefficient tensor.
The diffusion coefficient $\bm{D}(t)$ is assumed to obey another stochastic process which is independent
of $\bm{Q}$.
Although eq~\eqref{langevin_equation_mesoscopic_fluctuating_diffusivity} is not
the same as eq~\eqref{langevin_equation_mesoscopic_transient_potential}, they
are similar in some aspects.
Both of them employ additional
degrees of freedom to describe the mesoscopic dynamics.
In addition, the LEFD model can reproduce the non-Gaussian
behavior successfully\cite{Uneyama-Miyaguchi-Akimoto-2015}.

It would be informative to discuss how the LETP and the LEFD can be related
and whether these models can be unified or not. If we employ the LEFD
to describe the diffusion
of a single particle in a supercooled liquid (the same system as considered
in Sec.~\ref{example_supercooled_liquid}), we have
\begin{equation}
 \label{lefd_supercooled_liquid}
 \frac{d\bm{r}(t)}{dt} = \sqrt{2 D(t)} \bm{W}(t) ,
\end{equation}
where $D(t)$ is a scalar fluctuating diffusion coefficient.
We assume that $D(t)$ obeys an equilibrium stochastic process and the
statistical average of $D(t)$ is independent of time.
Then the MSD becomes
\begin{equation}
 \label{lefd_msd_supercooled_liquid}
 \langle [\bm{r}(t) - \bm{r}(0)]^{2} \rangle = 6 \langle D \rangle t.
\end{equation}
Eq~\eqref{lefd_msd_supercooled_liquid} means that the MSD of
the LEFD is simply proportional to $t$ for any $t$.
Therefore, unlike the GLEG and the LETP (eqs~\eqref{gleg_msd_supercooled_liquid}
and \eqref{msd_expression_supercooled_final}), the LEFD cannot describe
the MSD of a supercooled liquids. However, the fluctuation
of the diffusion coefficient strongly affects the higher order
correlation functions, unlike the GLEG.
Thus physical quantities which incorporate the higher order
correlation functions, such as the NGP, exhibit nontrivial behavior.
The NGP can be related to the correlation function of the fluctuating diffusivity as\cite{Uneyama-Miyaguchi-Akimoto-2015}
\begin{equation}
 \label{lefd_ngp_supercooled_liquid}
   \alpha(t)  = \frac{2}{t^{2}} \int_{0}^{t} dt' \int_{0}^{t'} dt'' 
 \left[ \frac{\langle D(t') D(t'') \rangle}{\langle D \rangle^{2}} - 1\right] 
    = \frac{2}{t^{2}} \int_{0}^{t} dt' \, (t - t')
 \left[ \frac{\langle D(t') D(0) \rangle}{\langle D \rangle^{2}} - 1\right].
\end{equation}
From eq~\eqref{lefd_ngp_supercooled_liquid}, in general, the LETP gives non-zero NGP, and therefore the
heterogeneity of the diffusion behavior can be successfully reproduced.
At the short time scale, eq~\eqref{lefd_ngp_supercooled_liquid} approximately
becomes independent of time: $\alpha(t) \approx \langle D^{2} \rangle / \langle D \rangle^{2} - 1$.
Generally, the NGP by eq~\eqref{lefd_ngp_supercooled_liquid}
becomes a monotonically decreasing function of time $t$.
Such behavior is qualitatively different from that of the LETP.
Therefore, we conclude that both the LETP and the LEFD can reproduce
non-Gaussian dynamics, but they are not equivalent.

We may interpret the LEFD as an approximation for the LETP
in the long region. If the time scale is larger than the average
relaxation time of the transient potential, we will observe simple
diffusion behavior where the MSD is approximately proportional to time.
Also, the NGP can be interpreted as a monotonically decreasing function of time.
These properties are qualitatively consistent with those of the LEFD.
Therefore, in such a case, the effect of the transient
potential on the dynamic equation may be further coarse-grained.
Then the thermodynamic force will be simply determined by the free energy,
and the LETP can be coarse-grained into the LEFD model.
It should be noted here that the GLEG cannot be employed for a system which exhibits non-Gaussian behavior.
As Fox showed\cite{Fox-1977}, the memory kernel is uniquely determined if the MSD
is given. At the long time scale, the memory kernel approximately becomes the delta function
and thus we just have a simple Langevin equation without memory effects
and the fluctuation of diffusivity.

\subsection{Transient Potential as Thermostat}
\label{transient_potential_as_thermostat}

One may consider the structure of the LETP is somewhat similar to some
thermostat models in molecular dynamics simulations. The Nos\'{e}-Hoover
thermostat utilizes the extended Hamiltonian where the extra degrees of
freedom for the thermostat are incorporated\cite{Evans-Holian-1985,Evans-Morris-book}.
Leimkuhler, Noorizadeh and Theil\cite{Leimkuhler-Noorizadeh-Theil-2009}
proposed a modified version of the Nos\'{e}-Hoover thermostat which
employs the Langevin equation for the dynamics of the thermostat.
We expect that the transient potential with the pseudo thermodynamic
degrees of freedom may work as a thermostat.
In this subsection, we consider a possible application of the transient
potential as a thermostat.

From eqs~\eqref{master_equation_approx}-\eqref{transition_rate_matrix_approx},
the approximate dynamics model for the coarse-grained system is detailed-balance.
Therefore, if we simply omit the noise term in the Langevin equation for $\bm{Q}$,
the resulting dynamics becomes physically incorrect, since the detailed-balance condition
is no longer satisfied. Therefore, we consider the Hamiltonian-like dynamics
for $\bm{Q}$.
We hypothetically introduce the momentum $\bm{P}$ and mass $m$, and
assume that the system obeys the following dynamic equations:
\begin{equation}
 \label{hamilton_equation_with_transient_potential}
 \frac{d\bm{P}(t)}{dt} = - \frac{\partial \check{\Phi}(\bm{Q},\bm{A})}{\partial \bm{Q}}, \qquad
 \frac{d\bm{Q}(t)}{dt} = \frac{1}{m} \bm{P}.
\end{equation}
Eq~\eqref{hamilton_equation_with_transient_potential} corresponds to the
Hamilton's canonical equations for the hypothetical Hamiltonian $\mathcal{H} = \bm{P}^{2} / 2 m + \check{\Phi}(\bm{Q},\bm{A})$.
We further assume that the transient potential is given as the sum of the effective
interaction potential $\bar{U}(\bm{Q})$ and the harmonic potential as
\begin{equation}
 \label{transient_potential_thermostat}
 \check{\Phi}(\bm{Q},\bm{A}) = \bar{U}(\bm{Q}) + \frac{\kappa}{2} (\bm{Q} - \bm{A})^{2},
\end{equation}
where $\kappa$ is a constant. The variables $\bm{Q}$ and $\bm{P}$ are coupled to
the stochastic variable $\bm{A}$ via the harmonic potential, and thus we expect that 
the equilibrium state will be realized.

To demonstrate the transient potential actually works as a thermostat, we consider 
the case where $\bm{A}$ obeys the overdamped Langevin equation \eqref{langevin_equation_mesoscopic_transient_potential_pseudo}.
If we assume that the mobility is given as $\bm{\Gamma} = \bm{1} / \zeta$ with $\zeta$ being the friction
coefficient, the dynamic equation becomes
\begin{equation}
 \label{langevin_equation_thermostat}
 \frac{d\bm{A}(t)}{dt} 
 % = - \frac{1}{\zeta} \frac{\partial \check{\Phi}(\bm{Q},\bm{A})}{\partial \bm{A}} + \sqrt{\frac{2 k_{B} T}{\zeta}} \bm{\omega}(t)
 = - \frac{\kappa}{\zeta} (\bm{A} - \bm{Q}) + \sqrt{\frac{2 k_{B} T}{\zeta}} \bm{\omega}(t).
\end{equation}
From eqs \eqref{hamilton_equation_with_transient_potential}
and \eqref{langevin_equation_thermostat}, we have
\begin{equation}
 \label{langevin_equation_thermostat_modified}
 \bm{A}(t)  = \bm{Q}(t) - \frac{1}{\kappa} \int_{-\infty}^{t} dt' \, K(t - t') \frac{1}{m} \bm{P}(t') + \bm{\xi}(t)
\end{equation}
with
$K(t - t') = \kappa e^{-t \kappa / \zeta}$ and
$\bm{\xi}(t) \equiv \sqrt{2 k_{B} T / \zeta} \int_{-\infty}^{t} dt' K(t - t') \bm{\omega}(t')$.
By substituting eq~\eqref{langevin_equation_thermostat_modified} into
eq~\eqref{hamilton_equation_with_transient_potential}, the dynamic equation for $\bm{Q}(t)$ can be
simply expressed as
\begin{equation}
 \label{hamilton_equation_with_transient_potential_modified}
 m \frac{d^{2}\bm{Q}(t)}{dt^{2}} = - \frac{\partial \bar{U}(\bm{Q}(t))}{\partial \bm{Q}(t)}
  - \int_{-\infty}^{t} dt' \, K(t - t') \frac{d\bm{Q}(t')}{dt'} + \bm{\xi}(t).
\end{equation}
The noise $\bm{\xi}(t)$ is a linear combination of the Gaussian white noise $\bm{\omega}(t)$ and becomes a Gaussian colored noise.
The first and second moments of $\bm{\xi}(t)$ are calculated to be
\begin{equation}
 \label{fluctuation_dissipation_relation_thermostat}
 \langle \bm{\xi}(t) \rangle = 0, \qquad
 \langle \bm{\xi}(t) \bm{\xi}(t') \rangle = k_{B} T K(|t - t'|) \bm{1}.
\end{equation}
Eq~\eqref{fluctuation_dissipation_relation_thermostat} can be interpreted as the fluctuation-dissipation relation.
Therefore we find that $\bm{Q}(t)$ obeys the GLEG with the memory kernel $K(t)$, and thus
the transient potential works as a thermostat.
Although we have not explicitly introduced the memory kernel in
eqs~\eqref{hamilton_equation_with_transient_potential}-\eqref{langevin_equation_thermostat},
the resulting dynamics reproduces the memory effect. If we employ a non-harmonic
transient potential model and/or a transition dynamics model, we will be able
to reproduce a non-Gaussian thermostat as well.

%------------------------------------------------------------------------------
\section{Conclusions}
\label{colcusions}

We showed that we can formally derive
the transient potential model (LETP) starting from the microscopic
Langevin equation model. We showed that we can formally justify the use of
the transient potential, based on the path probability formalism which
utilizes the Onsager-Machlup action.
However, the dynamics for the transient potential is generally
not given in a simple and tractable form. Instead of the exact dynamics
for the transient potential, we proposed to introduce
the pseudo thermodynamic degrees of freedom and employ simple approximate
dynamics model.
The obtained LETP consist of two dynamics models;
one is the simple Langevin equation for the mesoscopic degrees of freedom,
and another is the Markovian stochastic dynamics model for
the additional degrees of freedom
(the pseudo thermodynamic degrees of freedom).
The LETP can reproduce
non-Gaussian dynamics which the GLEG cannot reproduce. As a simple example,
we considered the dynamics of a tagged particle in a supercooled liquid.
We found that the LETP can qualitatively reproduce the characteristic diffusion
behavior.

We expect that the LETP can be utilized as a general coarse-grained
equation for mesoscopic dynamics of soft matters. The result of this
work justifies the mesoscopic dynamics model such as the RaPiD and
MCSS model which were originally introduced as purely phenomenological
models. However, at least currently, the derivation of the LETP is limited
to rather simple systems.
The underlying microscopic dynamics model is
assumed to be the overdamped Langevin equation with the constant mobility
tensor. The mesoscopic degrees of freedom are limited to the linear combinations
of microscopic degrees of freedom. More general derivations and detailed
analyses will be required to
further elaborate the coarse-grained dynamics models.
For example, the derivation of the LETP from the microscopic Hamiltonian dynamics
is an interesting future work.
In addition, the development of
accurate and practical approximation models for the transient potential is
also required. Although we simply assumed the Markovian process for the
pseudo thermodynamic degrees of freedom in this work, other dynamics models
would be employed instead.

%------------------------------------------------------------------------------
\section*{Acknowledgment}
\label{acknowledgment}

This work was supported by Grant-in-Aid (KAKENHI) for Scientific Research Grant C
No. JP16K05513 from Ministry of Education, Culture, Sports, Science, and Technology,
and Grant-in-Aid (KAKENHI) for Scientific Research Grant B No. JP19H01861
from Ministry of Education, Culture, Sports, Science, and Technology,
and JST, PRESTO Grant Number JPMJPR1992.

%------------------------------------------------------------------------------
% appendix
\appendix

\section{Multiplicative Noise and Nonlinear Variable Transform}
\label{multipliticative_noise_and_nonlinear_variable_transform}

In this appendix, we consider the coarse-graining for a system
described by the overdamped Langevin equation with the multiplicative noise.
We employ the following Langevin equation with the position-dependent mobility
as the microscopic dynamic equation, instead of eq~\eqref{langevin_equation_microscopic_modified}:
\begin{equation}
 \label{langevin_equation_microscopic_multiplicative}
 \frac{d\bm{R}(t)}{dt} = -\bm{L}(\bm{R}) \cdot \frac{\partial U(\bm{R})}{\partial \bm{R}}
  + k_{B} T \frac{\partial}{\partial \bm{R}} \cdot \bm{L}(\bm{R})
  + \sqrt{2 k_{B} T} \bm{L}^{1/2}(\bm{R}) \cdot \bm{w}(t) ,
\end{equation}
where $\bm{L}(\bm{R})$ is the position-dependent mobility. The noise term
in eq~\eqref{langevin_equation_microscopic_multiplicative} is multiplicative and
we interpret it according to the Ito manner.

As the same way in the main text, we introduce 
the variable transform from $\bm{R}$ to $\bm{X} \equiv [\bm{Q}^{\mathrm{T}} \, \bm{\theta}^{\mathrm{T}}]^{\mathrm{T}}$
($\bm{Q}$ is an $M$-dimensional vector and $\bm{\theta}$ is a $(3 N - M)$-dimensional vector).
This transform can be nonlinear, but the inverse transform should exist.
$\bm{X}$ can be interpreted as a function of $\bm{R}$, as
$\bm{X}(\bm{R})$.
The inverse transform exists if the following condition is satisfied:
\begin{equation}
 \det \frac{\partial \bm{X}}{\partial \bm{R}} \neq 0,
\end{equation}
where $\partial \bm{X} / \partial \bm{R}$ corresponds to the Jacobian matrix
for the variable transform.
Then, $\bm{R}$ can be interpreted as the function of $\bm{X}$,
as $\bm{R}(\bm{X})$.
The effective interaction potential for $\bm{X}$ becomes\cite{Nakamura-2018}
\begin{equation}
 \label{effective_interaction_potential_multiplicative}
 U'(\bm{X}) = U(\bm{R}(\bm{X})) + k_{B} T \ln \det \frac{\partial \bm{X}}{\partial \bm{R}}.
\end{equation}
The second term in the right hand side of eq~\eqref{effective_interaction_potential_multiplicative}
arises from the metric of the nonlinear variable transform.
If the variable transform is linear and $\bm{X}$ is linear in $\bm{R}$
(as the case we considered in the main text),
it reduces to a constant and negligible. The mobility tensor becomes\cite{Uneyama-2020}
\begin{equation}
 \label{mobility_tensor_multiplicative}
 \bm{L}'(\bm{X}) = 
  \begin{bmatrix}
   {L'}_{ij}^{(\bm{Q})}(\bm{X}) & {L'}_{i\beta}^{(\bm{Q\theta})}(\bm{X}) \\
   {L'}^{(\bm{\theta\bm{Q}})}_{\alpha j}(\bm{X}) & {L'}_{\alpha\beta}^{(\bm{\theta})}(\bm{X})
  \end{bmatrix},
\end{equation}
with
\begin{align}
 {L'}_{ij}^{(\bm{Q})}(\bm{X}) & = \frac{\partial Q_{i}(\bm{r})}{\partial \bm{r}} \cdot \bm{L}(\bm{R}) \cdot \frac{\partial Q_{j}(\bm{R})}{\partial \bm{R}}, \\
 {L'}_{i\beta}^{(\bm{Q\theta})}(\bm{X}) & = {L'}_{\alpha j}^{(\bm{\theta\bm{Q}})}(\bm{X}) = \frac{\partial Q_{i}(\bm{R})}{\partial \bm{R}} \cdot \bm{L}(\bm{R}) \cdot \frac{\partial \theta_{\alpha}(\bm{R})}{\partial \bm{R}}, \\
 {L'}_{\alpha\beta}^{(\bm{\theta})}(\bm{X}) & =  \frac{\partial \theta_{\alpha}(\bm{R})}{\partial \bm{R}} \cdot \bm{L}(\bm{r}) \cdot \frac{\partial \theta_{\beta}(\bm{R})}{\partial \bm{R}}.
\end{align}

So far, any $\bm{\theta}$ can be employed as long as the variable transform is
invertible. Here we employ $\bm{\theta}$ which is not kinetically coupled
to $\bm{Q}$. That is, we employ $\bm{\theta}$ which
satisfies the following condition:
\begin{equation}
 \label{orthogonal_condition_multiplicative}
  \frac{\partial Q_{i}(\bm{R})}{\partial \bm{R}} \cdot \bm{L}(\bm{R}) \cdot \frac{\partial \theta_{\alpha}(\bm{R})}{\partial \bm{R}} = 0.
\end{equation}
Then the mobility tensor \eqref{mobility_tensor_multiplicative} becomes block-diagonal:
\begin{equation}
 \bm{L}'(\bm{X}) = 
  \begin{bmatrix}
   {L'}_{ij}^{(\bm{Q})}(\bm{X}) & 0 \\
   0 & {L'}_{\alpha\beta}^{(\bm{\theta})}(\bm{X})
  \end{bmatrix} .
\end{equation}
The problem is that whether such $\bm{\theta}$ actually exists or not.
Fortunately, we can show that we can construct $\bm{\theta}$ which satisfies
eq~\eqref{orthogonal_condition_multiplicative}
for any $\bm{Q}$. Eq~\eqref{orthogonal_condition_multiplicative} 
can be rewritten as
\begin{equation}
 \label{orthogonal_condition_multiplicative_modified}
 \bm{u}_{i}(\bm{R}) \cdot \frac{\partial \theta_{\alpha}(\bm{R})}{\partial \bm{R}} = 0,
\end{equation}
with 
$\bm{u}_{i}(\bm{R}) \equiv [\partial Q_{i}(\bm{R})/\partial \bm{R}] \cdot \bm{L}(\bm{R})$  ($i = 1,2,\dots,M$).
%\begin{equation}
%  \bm{u}_{i}(\bm{R}) \equiv  \frac{\partial Q_{i}(\bm{R})}{\partial \bm{R}} \cdot \bm{L}(\bm{R}).
%\end{equation}
Here $\bm{u}_{i}(\bm{R})$ is a $3N$-dimensional vector. This $\bm{u}_{i}(\bm{R})$
can be expanded into the position-dependent orthogonal basis $\bm{e}^{\parallel}_{i}(\bm{R})$ ($i = 1,2,\dots,M$), as
\begin{equation}
 \bm{u}_{i}(\bm{R}) = \sum_{j}
  \left[ \bm{u}_{i}(\bm{R}) \cdot \bm{e}^{\parallel}_{j}(\bm{R}) \right] \bm{e}^{\parallel}_{j}(\bm{R}) .
\end{equation}
The position vector $\bm{R}$ is a $3 N$-dimensional vector, thus we can construct
$(3 N - M)$ orthogonal basis vectors which are orthogonal to $\bm{e}^{\parallel}_{i}(\bm{R})$.
If we describe this basis as $\bm{e}^{\perp}_{\alpha}(\bm{R})$ ($\alpha = M + 1, M + 2, \dots, 3 N$),
we simply have
$\bm{e}^{\parallel}_{i}(\bm{R}) \cdot \bm{e}^{\perp}_{\alpha}(\bm{R}) = 0$.
%\begin{equation}
%  \bm{e}^{\parallel}_{i}(\bm{R}) \cdot \bm{e}^{\perp}_{\alpha}(\bm{R}) = 0.
%\end{equation}
This means that the condition \eqref{orthogonal_condition_multiplicative_modified} can be satisfied if
we take $\bm{\theta}$ which satisfies the following condition:
\begin{equation}
 \label{orthogonal_condition_multiplicative_modified2}
 \frac{\partial \theta_{\alpha}(\bm{R})}{\partial \bm{R}} = \rho_{\alpha} \bm{e}^{\perp}_{\alpha}(\bm{R}),
\end{equation}
where $\rho_{\alpha}$ is constant. (Notice that we do not take the summation over $\alpha$ in the
right hand side of \eqref{orthogonal_condition_multiplicative_modified2}.)
We may further rewrite eq~\eqref{orthogonal_condition_multiplicative_modified2} as
\begin{equation}
 \label{orthogonal_condition_multiplicative_modified3}
 \frac{\partial^{2} \theta_{\alpha}(\bm{R})}{\partial \bm{R}^{2}} =
 \rho_{\alpha} \frac{\partial}{\partial \bm{R}} \cdot \bm{e}^{\perp}_{\alpha}(\bm{R}).
\end{equation}
Eq~\eqref{orthogonal_condition_multiplicative_modified3}
is a Poisson equation in the $3 N$-dimensional space. The solution is
\begin{equation}
 \label{theta_constructed_multiplicative}
 \theta_{\alpha}(\bm{R}) = \bar{\theta}_{\alpha} +
 \rho_{\alpha} \bar{\bm{e}}^{\perp}_{\alpha} \cdot \bm{R}
  + \rho_{\alpha} \int d\bm{R}' \, G(\bm{R} - \bm{R}') \frac{\partial}{\partial \bm{R}'} \cdot 
  \left[ \bm{e}^{\perp}_{\alpha}(\bm{R}') - \bar{\bm{e}}^{\perp}_{\alpha} \right],
\end{equation}
where $\bar{\theta}_{\alpha}$ is a constant, $\bar{\bm{e}}^{\perp}_{\alpha}$ is the spatial average of $\bm{e}_{\alpha}^{\perp}(\bm{R})$,
 and $G(\bm{R})$ is the Green function for the Poisson equation:
\begin{equation}
 - \frac{\partial G(\bm{R})}{\partial \bm{R}^{2}} = \delta(\bm{R}).
\end{equation}
In the three dimensional space, the Green function becomes a simple Coulomb type kernel.
(In a $3 N$-dimensional space ($3 N \ge 3$), the Green function $G(\bm{R})$ decays as $|\bm{R}|^{2 - 3 N}$ for large $|\bm{R}|$.)
$\bm{\theta}(\bm{R})$ given by \eqref{theta_constructed_multiplicative}
satisfies eq~\eqref{orthogonal_condition_multiplicative}, and
the mobility tensor can be block-diagonal.
We should notice that the basis vector $\bm{e}_{\alpha}^{\perp}(\bm{R})$ depends on $\bm{Q}$ and
thus is not constant.
To satisfy eq~\eqref{orthogonal_condition_multiplicative}
for any $t$, we should modulate $\bm{\theta}(t)$ during the time evolution.
This can be done by introducing the Lagrange multiplier into the Langevin
equation for $\bm{\theta}(t)$. (Intuitively, the Lagrange multiplier
can be understood as the external force which drives $\bm{\theta}(t)$ to satisfy
the condition \eqref{orthogonal_condition_multiplicative_modified}.)

The Onsager-Machlup action and the path probability becomes
\begin{equation}
 \exp[-\mathcal{S}[\bm{R}]] \mathcal{D}\bm{R}
  = \exp[-\mathcal{S}[\bm{Q},\bm{\theta}]]
  \mathrm{Det} \frac{\delta \bm{X}}{\delta \bm{R}}
  \mathcal{D}\bm{Q}\mathcal{D}\bm{\theta},
\end{equation}
where $\mathrm{Det}\dotsb$ represents the functional determinant, and the
action $\mathcal{S}[\bm{Q},\bm{\theta}]$ is given as follows:
\begin{equation}
 \mathcal{S}[\bm{Q},\bm{\theta}]
  = \mathcal{S}^{(\bm{Q})}[\bm{Q}|\bm{\theta}]
 + \mathcal{S}^{(\bm{\theta})}[\bm{\theta}|\bm{Q}] ,
\end{equation}
\begin{equation}
 \label{effective_action_multiplicative_q}
  \mathcal{S}^{(\bm{Q})}[\bm{Q}|\bm{\theta}]
   = \frac{1}{2 k_{B} T} \int dt \,
  G\left( \frac{d\bm{Q}}{dt} + {\bm{L}'}^{(\bm{Q})} \cdot \frac{\partial U'}{\partial \bm{Q}}
    - k_{B} T \frac{\partial }{\partial \bm{Q}} \cdot {\bm{L}'}^{(\bm{Q})}
 ; {\bm{L}'}^{(\bm{Q})} \right) ,
\end{equation}
\begin{equation}
 \label{effective_action_multiplicative_theta}
 \mathcal{S}^{(\bm{\theta})}[\bm{\theta}|\bm{Q}]
  = \frac{1}{2 k_{B} T} \int dt \,
  G\left( \frac{d\bm{\theta}}{dt} - \bm{\Upsilon} + {\bm{L}'}^{(\bm{\theta})} \cdot \frac{\partial U'}{\partial \bm{\theta}}
    - k_{B} T \frac{\partial }{\partial \bm{\theta}} \cdot {\bm{L}'}^{(\bm{\theta})}
 ; {\bm{L}'}^{(\bm{\theta})} \right)  .
\end{equation}
Here, $\bm{\Upsilon}(t)$ is the time-dependent Lagrange multiplier for
the condition \eqref{orthogonal_condition_multiplicative}.
Now the situation is similar to that in the main text.
We introduce the transient potential $\Phi(\tilde{\bm{q}},t)$ by the functional
identity~\eqref{delta_functional_identity}.
Also, we introduce the time-dependent and fluctuating mobility (diffusivity)\cite{Uneyama-Miyaguchi-Akimoto-2015,Miyaguchi-Akimoto-Yamamoto-2016,Miyaguchi-2017,Uneyama-Miyaguchi-Akimoto-2019,Miyaguchi-Uneyama-Akimoto-2019} by
utilizing another functional identity:
\begin{equation}
 \label{delta_functional_identity_multiplicative}
 1 = \int \mathcal{D}\bm{\Lambda} \, \delta[\bm{\Lambda}(\tilde{\bm{q}},t) - {\bm{L}'}^{(\bm{Q})}(\tilde{\bm{q}},\bm{\theta}(t))] .
\end{equation}
By utilizing eqs~\eqref{delta_functional_identity} and \eqref{delta_functional_identity_multiplicative}, we can
rewrite the path probability as
\begin{equation}
\begin{split}
 \label{effective_path_probability_multiplicative}
 \mathcal{P}[\bm{Q}]
 &  = \int\mathcal{D}\bm{\theta} \mathcal{D}\Phi \mathcal{D}\bm{\Lambda} \,
  \frac{\mathcal{N}^{(\bm{Q},\bm{\theta},\Phi,\bm{\Lambda})}}{\sqrt{\mathrm{Det} {\bm{L}'}^{(\bm{Q})} \, \mathrm{Det} {\bm{L}'}^{(\bm{\theta})}}}
  \exp\left[-\mathcal{S}^{(\bm{Q})}[\bm{Q}|\bm{\theta}]
       -\mathcal{S}^{(\bm{\theta})}[\bm{\theta}|\bm{Q}]
      \right] \\
 & \qquad \times   \mathrm{Det} \frac{\delta \bm{X}}{\delta \bm{R}}
\delta[\Phi(\tilde{\bm{q}},t) - U'(\tilde{\bm{q}},\bm{\theta}(t))] 
  \delta[\bm{\Lambda}(\tilde{\bm{q}},t) - {\bm{L}'}^{(\bm{Q})}(\tilde{\bm{q}},\bm{\theta}(t))] \\
 &  = \int \mathcal{D}\Phi \mathcal{D}\bm{\Lambda} \,
  \frac{\mathcal{N}^{(\bm{Q},\Phi,\bm{\Lambda})}}{\sqrt{\mathrm{Det} \bm{\Lambda}}}
  \exp\left[-\mathcal{S}^{(\bm{Q})}[\bm{Q}|\Phi,\bm{\Lambda}]
 - \mathcal{S}^{(\Phi,\bm{\Lambda})}[\Phi,\bm{\Lambda}|\bm{Q}]
      \right] ,
\end{split}
\end{equation}
with
\begin{equation}
 \tilde{\mathcal{S}}^{(\bm{Q})}[\bm{Q}|\Phi,\bm{\Lambda}]
  = \frac{1}{2 k_{B} T} \int dt \,
  G\left(\frac{d \bm{Q}}{dt} + \bm{\Lambda} \cdot \frac{\partial \Phi}{\partial \bm{Q}}
    - k_{B} T \frac{\partial}{\partial \bm{Q}} \cdot \bm{\Lambda}
   ;  \bm{\Lambda}
      \right) , 
\end{equation}
\begin{equation}
\begin{split}
   \tilde{\mathcal{S}}^{(\Phi,\bm{\Lambda})}[\Phi,\bm{\Lambda}|\bm{Q}]
 & = - \ln \int\mathcal{D}\bm{\theta} \,
  \frac{  \exp\left[ -\mathcal{S}^{(\bm{\theta})}[\bm{\theta}|\bm{Q}] \right]}{\sqrt{\mathrm{Det} {\bm{L}'}^{(\bm{\theta})}}}   \mathrm{Det} \frac{\delta \bm{X}}{\delta \bm{R}} \\
 & \qquad \times \delta[\Phi(\tilde{\bm{q}},t) - U'(\tilde{\bm{q}},\bm{\theta}(t))] 
  \delta[\bm{\Lambda}(\tilde{\bm{q}},t) - {\bm{L}'}^{(\bm{Q})}(\tilde{\bm{q}},\bm{\theta}(t))] .
\end{split}  
\end{equation}

Finally we have the following Langevin equation for $\bm{Q}(t)$:
\begin{equation}
 \label{langevin_equation_mesoscopic_transient_potential_fluctuating_diffusivity}
 \frac{d\bm{Q}(t)}{dt}
  = - \bm{\Lambda}(\bm{Q},t) \cdot \frac{\partial \Phi(\bm{Q},t)}{\partial \bm{Q}}
  + k_{B} T \frac{\partial }{\partial \bm{Q}} \cdot \bm{\Lambda}(\bm{Q},t)
  + \sqrt{2 k_{B} T} \bm{\Lambda}^{1/2}(\bm{Q},t) \cdot \bm{W}(t) ,
\end{equation}
where $\bm{W}(t)$ is the Gaussian white noise which satisfies eq~\eqref{fluctuation_dissipation_mesoscopic_transient_potential}.
Eq~\eqref{langevin_equation_mesoscopic_transient_potential_fluctuating_diffusivity}
has the same form as the LETP \eqref{langevin_equation_mesoscopic_transient_potential}.
However, in addition to the transient potential $\Phi(\bm{Q},t)$, the fluctuating 
mobility $\bm{\Lambda}(\bm{Q},t)$ is also incorporated in eq~\eqref{langevin_equation_mesoscopic_transient_potential_fluctuating_diffusivity}.
Therefore, for the systems with multiplicative noises and/or coarse-grained variables by nonlinear transforms,
we have the Langevin equation with two transient and fluctuating quantities;
the transient potential and the fluctuating mobility (diffusivity).
If the mobility tensor for $\bm{R}$ is constant and the variable transform from $\bm{R}$ to
$\bm{X}$ is linear, then $\bm{\Lambda}(\bm{Q},t)$ reduces to a constant and the LETP is recovered.

\section{Molecular Dynamics Simulation for Supercooled Liquid}
\label{molecular_dynamics_simulation_for_supercooled_liquid}

In this appendix, we show the details of the molecular dynamics
simulation model for a supercooled liquid used in the main text.
We employ a binary Lennard-Jones mixture type model\cite{Kob-Andersen-1994,Kob-Andersen-1995,Kob-Andersen-1995a,Yamamoto-Onuki-1998,Yamamoto-Onuki-1998a,Kob-1999,Vorselaars-Lyulin-Karatasos-Michels-2007}.
In this model,
we consider two particle species, A and B.
To prevent the crystallization, 
the A and B particles have different sizes $\sigma_{\text{A}}$ and $\sigma_{\text{B}}$.
The ratios of sizes and masses are set as $\sigma_{\text{B}} / \sigma_{\text{A}} = 1.2$ and
as $m_{\text{B}} / m_{\text{A}} = 2$, respectively, and
the number fraction of the A particles is $1/2$.
The interaction potential between particle species $K$ and $K'$
is given as the Lennard-Jones type potential:
\begin{equation}
 \label{repulsive_lennard_jones_potential}
 u_{KK'}(\bm{r}) = 
  \begin{cases}
  4 \varepsilon 
 [ ( {\sigma_{KK'}}/{|\bm{r}|})^{12}
   - ( {\sigma_{KK'}} / {|\bm{r}|})^{6}
   + 1 / 4  ] & (|\bm{r}| < 2^{1/6}\sigma_{KK'}) ,\\
   0 & (|\bm{r}| \ge 2^{1/6}\sigma_{KK'}) ,
  \end{cases}
\end{equation}
where $\sigma_{KK'} \equiv (\sigma_{K} + \sigma_{K'}) / 2$ and $\varepsilon$ is
the Lennard-Jones potential parameter. In eq~\eqref{repulsive_lennard_jones_potential}
We have truncated the Lennard-Jones
potential so that the potential becomes purely repulsive.

We consider a three dimensional system which consists of $N$ particles.
We use a cubic simulation box of which volume is $L^{3}$, and use the
periodic boundary condition.
We express the position of the $i$-th particle in the system as $\bm{r}_{i}$.
The particle species is A for $i = 1,2,\dots, N  /2$ and B for $i = N/2 + 1,N/2 + 2,\dots,N$.
The total potential energy of the system simply becomes
\begin{equation}
 U(\lbrace \bm{r}_{i} \rbrace) = 
  \sum_{i = 1}^{N/2}\sum_{j = 1}^{i} u_{\text{AA}}(\bm{r}_{i} - \bm{r}_{j})
  + \sum_{i = N/2 + 1}^{N}\sum_{j = 1}^{i} u_{\text{BB}}(\bm{r}_{i} - \bm{r}_{j})
  + \sum_{i = N}^{N/2}\sum_{j = N/2 + 1}^{N} u_{\text{AB}}(\bm{r}_{i} - \bm{r}_{j}) .
\end{equation}
As the dynamic equation, we employ the underdamped Langevin equation:
\begin{equation}
 m_{i} \frac{d\bm{r}_{i}(t)}{dt}
  = - \frac{\partial U(\lbrace \bm{r}_{i}(t) \rbrace)}{\partial \bm{r}_{i}(t)}
  - \zeta \frac{d\bm{r}_{i}(t)}{dt}
  + \sqrt{2 k_{B} T \zeta} \bm{w}_{i}(t) ,
\end{equation}
where $m_{i}$ is the mass of the $i$-th particle, $\zeta$ is the friction
coefficient, $\bm{w}_{i}(t)$ is the Gaussian white noise which satisfies
the fluctuation-dissipation relation.

To perform simulations, we employ usual Lennard-Jones dimensionless units
by setting $\sigma = \sigma_{\text{A}} = 1$, $ m = m_{\text{A}} = 1$, and $\varepsilon = 1$.
In this work, we set $N = 4000$ and $L = 17.1$ (this gives the average
number density as $\rho = N / L^{3} = 0.800$). The friction coefficient
is set as $\zeta = 10$. The characteristic momentum relaxation time is estimated
to be $\tau_{m} = m / \zeta = 0.1$.
Initially, the particles are randomly
placed in the box and then relaxed before the simulation starts.
Simulations are performed for different temperatures ranging from $k_{B} T = 0.4$
to $k_{B} T = 1$. The time step size is $\Delta t = 2.0 \times 10^{-3}$ and
simulations are performed for $t = 10^{5}$ for each temperature.
To remove the artificial diffusion behavior due to the center of mass
motion of the system, the momentum of the system is set to zero at
each time step.
All the simulations are performed with LAMMPS
(22Aug18)\cite{Plimpton-1995,lammps-web}.
The particle trajectories are recorded and then the MSD
and NGP are calculated.
To improve the statistical accuracy, several runs with the same
parameter set and the different initial structures and random seeds are
performed, and then the averages are taken over different runs.

%------------------------------------------------------------------------------
\section{Detailed Calculations for MSD and NGP} 
\label{detailed_calculation_for_msd_and_ngp}

The LETP model for a tagged particle in a supercooled liquid in the main text
consists of two stochastic processes (which are characterized by
eqs~\eqref{langevin_equation_transient_potential_supercooled}
and \eqref{transition_rate_transient_potential_supercooled}); one is the Langevin equation for
the particle and another is the resampling process for the potential center.
The Langevin equation describes the continuum process whereas the resampling
process is discrete in time.
We utilize the renewal
theory\cite{Godreche-Luck-2001}
which is suitable for the analyses of the resampling type process.
The analyses shown in this appendix are based on those in Ref.~\cite{Hachiya-Uneyama-Kaneko-Akimoto-2019}.

We consider the statistics of the resampling events
from time $0$. We describe the $i$-th resampling event occurs at time $t_{i}$.
For convenience, we set $t_{0} = 0$. We call
the interval between two successive resamplings as the waiting time.
During the time between successive resamplings, the potential center position
does not change. We express
the potential center for
$t_{i} < t < t_{i + 1}$ as $\bm{A}_{i}$. Also, we express $\bm{r}_{i} = \bm{r}(t_{i})$.
Without loss of generality, we can set the initial position of the particle
as $\bm{r}(0) = \bm{r}_{0} = 0$.
Since the resampling events are statistically independent, the interval
between two successive resamplings (the waiting time) is given as the
exponential distribution:
\begin{equation}
 \label{waiting_time_distribution}
 \Psi(t_{i + 1} - t_{i}) = \frac{1}{\tau} e^{- (t_{i + 1} - t_{i}) / \tau} .
\end{equation}
Here, $\tau$ is the characteristic time of the transition in eq~\eqref{transition_rate_transient_potential_supercooled},
and can be interpreted as the average waiting time.
The statistical properties of the displacement
can be calculated by using the probability distribution of the particle
position at time $t$, $P(\bm{r};t)$.

For $t_{i} < t' < t < t_{i + 1}$, no resampling occurs and the Langevin equation for
$\bm{r}$ reduces to the Ornstein-Uhlenbeck process\cite{vanKampen-book}. Thus the propagator can be easily calculated:
%\begin{equation}
% Q(\bm{r},t|\bm{r}',\bm{A}_{i},t') = 
%  \left[\frac{\kappa}{2 \pi (1 - e^{-2 \Lambda \kappa (t - t')}) k_{B} T}\right]^{3/2}
%  \exp\left[ - \frac{\kappa [(\bm{r} - \bm{A}_{i}) - e^{-\Lambda \kappa (t  -t')}(\bm{r}' - \bm{A}_{i})]^{2}}{2 (1 - e^{-2 \Lambda \kappa (t - t')}) k_{B} T }\right] .
%\end{equation}
\begin{equation}
 \label{propagator_ornstein_uhlenbeck}
 Q(\bm{r},t|\bm{r}',\bm{A}_{i},t') = 
  \left[\frac{\kappa}{2 \pi (1 - e^{-2 \Lambda \kappa (t - t')}) k_{B} T}\right]^{3/2}
  \exp\left[ - \frac{\kappa 
       [(\bm{r} - \bm{r}')
       - (1 - e^{-\Lambda \kappa (t  -t')}) (\bm{r}' - \bm{A}_{i})]^{2}}{2 (1 - e^{-2 \Lambda \kappa (t - t')}) k_{B} T }\right] ,
\end{equation}
where $\bm{r}'$ represents the position at time $t'$.
At time $t_{i}$, the potential center is resampled from the equilibrium
distribution:
\begin{equation}
 \label{resampling_distribution}
  \Psi'(\bm{A}_{i},\bm{r}_{i}) =
  \left(\frac{\kappa}{2 \pi k_{B} T}\right)^{3/2}
  \exp\left[ - \frac{\kappa (\bm{r}_{i} - \bm{A}_{i})^{2}}{2 k_{B} T }\right] .
\end{equation}

We describe the number of total resampling events from time $0$ to time $t$ is $n$,
and calculate the probability distribution of the particle position at time $t$
for a given $n$, $P_{n}(\bm{r};t)$, by using eqs \eqref{waiting_time_distribution}-\eqref{resampling_distribution}.
The probability can be calculated as the product of propagates of the successive events.
The resampling times should satisfy $0 = t_{0} \le t_{1} \le t_{2} \le \dots t_{n} \le t$.
Thus we have
\begin{equation}
 \label{displacement_distribution_n}
 \begin{split}
  P_{n}(\bm{r};t)
  & = \int_{t}^{\infty} dt'
  \int_{0}^{t} dt_{n}
  \int_{0}^{t_{n}} dt_{n -1} \dots
  \int_{0}^{t_{2}} dt_{1} \int d\bm{A}_{n} d\bm{A}_{n - 1} \dots d\bm{A}_{0} \\
  & \qquad \times \int d\bm{r}_{n} d\bm{r}_{n - 1} \dots d\bm{r}_{1} \, Q(\bm{r},t | \bm{r}_{n},\bm{A}_{n},t_{n}) 
  \Psi(t' - t_{n}) \Psi'(\bm{A}_{n},\bm{r}_{n}) \\
  & \qquad \times
  \left[ \prod_{i = 1}^{n} 
  Q(\bm{r}_{i},t_{i} | \bm{r}_{i - 1},\bm{A}_{i - 1},t_{i - 1}) 
  \Psi(t_{i} - t_{i - 1}) \Psi'(\bm{A}_{i - 1},\bm{r}_{i - 1}) \right]  .
 \end{split}
\end{equation}
The integral over $t'$ in eq~\eqref{displacement_distribution_n}
can be easily calculated: $\int_{t}^{\infty} dt' \, \Psi(t' - t_{n}) = \tau \Psi(t - t_{n})$.
Also, the integral over $\bm{A}_{i}$ eq~\eqref{displacement_distribution_n} can be calculated straightforwardly:
\begin{equation}
 \label{average_propagator_ornstein_uhlenbeck}
 \begin{split}
  & \int d\bm{A}_{i - 1} \,
  Q(\bm{r}_{i},t_{i} | \bm{r}_{i - 1},\bm{A}_{i - 1},t_{i - 1}) 
    \Psi(t_{i} - t_{i - 1}) \Psi'(\bm{A}_{i - 1},\bm{r}_{i - 1})   \\
  & = \left[\frac{\kappa^{2}}{(1 - e^{-2 \Lambda \kappa (t_{i - 1} - t_{i})}) (2 \pi k_{B} T)^{2}}\right]^{3/2}
  \int d\bm{A}_{i - 1} \,
  \exp\bigg[ - \frac{\kappa (\bm{r}_{i - 1} - \bm{A}_{i - 1})^{2}}{2 k_{B} T } \\
  & \qquad - \frac{\kappa 
       [(\bm{r}_{i} - \bm{r}_{i - 1})
       - (1 - e^{-\Lambda \kappa (t_{i - 1}  - t_{i})}) (\bm{r}_{i - 1} - \bm{A}_{i - 1})]^{2}}{2 (1 - e^{-2 \Lambda \kappa (t_{i - 1} - t_{i})}) k_{B} T } \bigg] 
    \Psi(t_{i} - t_{i - 1}) \\
%  & =  \left[\frac{\kappa^{2}}{(1 - e^{-2 \Lambda \kappa (t_{i - 1} - t_{i})}) (2 \pi k_{B} T)^{2}}\right]^{3/2}
%  \int d\bm{A}_{i - 1} \,
%  \exp\bigg[ - \frac{\kappa (\bm{r}_{i - 1} - \bm{A}_{i - 1})^{2}}{2 k_{B} T } \\
%  & \qquad - \frac{\kappa 
%       [(1 - e^{-\Lambda \kappa (t_{i - 1}  - t_{i})})^{-1} (\bm{r}_{i} - \bm{r}_{i - 1})
%       - (\bm{r}_{i - 1} - \bm{A}_{i - 1})]^{2}}{2 
%  \frac{(1 - e^{-2 \Lambda \kappa (t_{i - 1} - t_{i})})}{(1 - e^{-\Lambda \kappa (t_{i - 1}  - t_{i})})^{2}} k_{B} T } \bigg] \\
%  & =  \left[\frac{\kappa^{2}}{(1 - e^{-2 \Lambda \kappa (t_{i - 1} - t_{i})}) (2 \pi k_{B} T)^{2}}\right]^{3/2}
%  \int d\bm{A}_{i - 1} \,
%  \exp\bigg[ - \frac{\kappa 
%       [(1 - e^{-\Lambda \kappa (t_{i - 1}  - t_{i})})^{-2} (\bm{r}_{i} - \bm{r}_{i - 1})^{2}}{2 
%  \left[ 1 + 
%  \frac{(1 - e^{-2 \Lambda \kappa (t_{i - 1} - t_{i})})}{(1 - e^{-\Lambda \kappa (t_{i - 1}  - t_{i})})^{2}} \right] k_{B} T } \bigg] \\
%  & =  \left[\frac{\kappa^{2}}{(1 - e^{-2 \Lambda \kappa (t_{i - 1} - t_{i})}) (2 \pi k_{B} T)^{2}}\right]^{3/2}
%  \int d\bm{A}_{i - 1} \,
%  \exp\bigg[ - \frac{\kappa 
%       [(\bm{r}_{i} - \bm{r}_{i - 1})^{2}}{2 
%  \left[ (1 - e^{-\Lambda \kappa (t_{i - 1}  - t_{i})})^{2}  + 
%  (1 - e^{-2 \Lambda \kappa (t_{i - 1} - t_{i})}) \right] k_{B} T } \bigg] \\
  & = \frac{1}{\tau} \left[\frac{\kappa}{4 \pi (1 - e^{-2 \Lambda \kappa (t_{i - 1} - t_{i})}) k_{B} T}\right]^{3/2}
  \exp\left[ - \frac{\kappa (\bm{r}_{i} - \bm{r}_{i - 1})^{2}}{
  4 ( 1 - e^{-\Lambda \kappa (t_{i - 1}  - t_{i})} ) k_{B} T } - \frac{t_{i + 1} - t_{i}}{\tau} \right] \\
  & \equiv \bar{\Psi}(\bm{r}_{i} - \bm{r}_{i - 1},t_{i} - t_{i - 1}) .
 \end{split}
\end{equation}
Thus the probability \eqref{displacement_distribution_n} can be rewritten as follows:
\begin{equation}
 \label{displacement_distribution_n_modified}
 \begin{split}
  P_{n}(\bm{r};t)
  & = \tau
  \int_{0}^{t} dt_{n}
  \int_{0}^{t_{n}} dt_{n -1} \dots
  \int_{0}^{t_{2}} dt_{1} 
  \int d\bm{r}_{n} d\bm{r}_{n - 1} \dots d\bm{r}_{1} \\
  & \qquad \times 
  \bar{\Psi}(\bm{r} - \bm{r}_{n},t - t_{n}) 
 \prod_{i = 1}^{n}   \bar{\Psi}(\bm{r}_{i} - \bm{r}_{i - 1},t_{i} - t_{i - 1})
  .
 \end{split}
\end{equation}
Because eq~\eqref{displacement_distribution_n_modified}
contains multiple convolutions over positions and times,
the Fourier-Laplace transform is convenient. The Fourier-Laplace
transform of eq~\eqref{displacement_distribution_n_modified} can be straightforwardly calculated as
\begin{equation}
 \label{displacement_distribution_n_fourier_laplace}
  \hat{P}_{n}(\bm{k};s)
   \equiv \int_{0}^{\infty} dt  \int d\bm{r} \, e^{- st - i \bm{k} \cdot \bm{r}} P_{n}(\bm{r};t) 
   = \tau \hat{\Psi}^{n + 1}(\bm{k},s) ,
\end{equation}
where
\begin{equation}
  \label{average_propagator_ornstein_uhlenbeck_fourier_laplace}
 \begin{split}
  \hat{\Psi}(\bm{k},s) 
  & \equiv
  \int_{0}^{\infty} dt  \int d\bm{r} \, e^{- st - i \bm{k} \cdot \bm{r}} \bar{\Psi}(\bm{r},t) \\
%  & = 
%  \int_{0}^{\infty} dt  \int d\bm{r} \,  \frac{1}{\tau} \left[\frac{\kappa}{4 \pi (1 - e^{-2 \Lambda \kappa t}) k_{B} T}\right]^{3/2} 
%  \exp\left[ - s t - i \bm{k} \cdot \bm{r} - \frac{\kappa \bm{r}^{2}}{
%  4 ( 1 - e^{-\Lambda \kappa t} ) k_{B} T } - \frac{t}{\tau} \right] \\
  & = 
  \int_{0}^{\infty} dt \,  \frac{1}{\tau} 
  \exp\left[ - (s + 1 / \tau) t  - \frac{ ( 1 - e^{-\Lambda \kappa t} ) k_{B} T  \bm{k}^{2}}{\kappa} \right] .
 \end{split}
\end{equation}

For small $\bm{k}^{2}$, we can expand 
eq~\eqref{average_propagator_ornstein_uhlenbeck_fourier_laplace}
into the power series of $\epsilon \equiv - \bm{k}^{2}$ as
\begin{equation}
  \label{average_propagator_ornstein_uhlenbeck_fourier_laplace_expanded}
   \hat{\Psi}(\bm{k},s) 
   = \hat{\Psi}_{0}(u) 
   + \hat{\Psi}_{1}(u) \epsilon
   + \hat{\Psi}_{2}(u) \epsilon^{2}
 + O(\epsilon^{3}) ,
\end{equation}
%\begin{equation}
%  \hat{\Psi}(\bm{k},s) 
%   = \hat{\Psi}_{0}(s) 
%   - \hat{\Psi}_{1}(s) \bm{k}^{2}
%   + \hat{\Psi}_{2}(s) |\bm{k}|^{4}
% + O(|\bm{k}|^{6}) ,
%\end{equation}
where we have defined $u \equiv \tau s$, and the explicit forms of the expansion coefficients become as follows, with $\eta \equiv \Lambda \kappa \tau$:
\begin{align}
  \label{average_propagator_ornstein_uhlenbeck_fourier_laplace_0th}
 \hat{\Psi}_{0}(u) & = \frac{1}{u + 1}  , \\
  \label{average_propagator_ornstein_uhlenbeck_fourier_laplace_1st}
 \hat{\Psi}_{1}(u) & = \frac{ k_{B} T }{\kappa}
  \left( \frac{1}{u + 1} - \frac{1}{u + 1 + \eta} \right) , \\
  \label{average_propagator_ornstein_uhlenbeck_fourier_laplace_2nd}
 \hat{\Psi}_{2}(s) & =  \frac{ (k_{B} T)^{2} }{2 \kappa^{2}}
  \left( \frac{1}{u + 1} 
  - \frac{2}{u + 1 + \eta}
  + \frac{1}{u + 1 + 2 \eta}
  \right) .
\end{align}
%\begin{align}
% \hat{\Psi}_{0}(s) & = \frac{1}{\tau (s + 1 / \tau)}  , \\
% \hat{\Psi}_{1}(s) & = \frac{ k_{B} T }{\tau \kappa}
%  \left( \frac{1}{s + 1 / \tau} - \frac{1}{s + 1 / \tau + \Lambda \kappa} \right) , \\
% \hat{\Psi}_{2}(s) & =  \frac{ (k_{B} T)^{2} }{2 \tau \kappa^{2}}
%  \left( \frac{1}{s + 1 / \tau} 
%  - \frac{2}{s + 1 / \tau + \Lambda \kappa}
%  + \frac{1}{s + 1 / \tau + 2 \Lambda \kappa}
%  \right) .
%\end{align}
The probability of the position $\bm{r}$ at time $t$ is given as the sum of
$P_{n}(\bm{r};t)$ for $n = 0,1,2,\dots$:
\begin{equation}
 \label{displacement_distribution}
 P(\bm{r};t) = \sum_{n = 0}^{\infty} P_{n}(\bm{r};t),
\end{equation}
and its Fourier-Laplace transform becomes
\begin{equation}
 \label{displacement_distribution_fourier_laplace}
 \hat{P}(\bm{k};s) \equiv \int_{0}^{\infty} dt \int d\bm{r} \, e^{-i \bm{k}\cdot\bm{r} - s t}
  P(\bm{r};t)
  = \tau \sum_{n = 0}^{\infty} \hat{\Psi}^{n + 1}(\bm{k};s)
  = \frac{\tau \hat{\Psi}(\bm{k};s)}{1 - \hat{\Psi}(\bm{k};s)} .
\end{equation}
By substituting eq~\eqref{average_propagator_ornstein_uhlenbeck_fourier_laplace_expanded}
into eq~\eqref{displacement_distribution_fourier_laplace}, 
the power series expansion of eq~\eqref{displacement_distribution_fourier_laplace}
becomes
%\begin{equation}
%  \hat{P}(\bm{k};s) 
%   = \frac{1}{s}
%  - \frac{\tau \hat{\Psi}_{1}(s)}{[1 - \hat{\Psi}_{0}(s)]^{2}} \bm{k}^{2}
%  + \left[ \frac{\tau \hat{\Psi}_{2}(s)}{[1 - \hat{\Psi}_{0}(s)]^{2}} 
%  - \frac{\tau \hat{\Psi}_{1}^{2}(s)}{[1 - \hat{\Psi}_{0}(s)]^{3}} 
%  \right] |\bm{k}|^{4}
%  + O(|\bm{k}|^{6}) .
%\end{equation}
\begin{equation}
 \label{displacement_distribution_fourier_laplace_expanded}
  \hat{P}(\bm{k};s) 
   = \frac{\tau}{u}
  + \frac{\tau \hat{\Psi}_{1}(u)}{[1 - \hat{\Psi}_{0}(u)]^{2}} \epsilon
  + \left[ \frac{\tau \hat{\Psi}_{2}(u)}{[1 - \hat{\Psi}_{0}(u)]^{2}} 
  + \frac{\tau \hat{\Psi}_{1}^{2}(u)}{[1 - \hat{\Psi}_{0}(u)]^{3}} 
  \right] \epsilon^{2}
  + O(\epsilon^{3}) .
\end{equation}
The Laplace transforms of the MSD and the mean-quartic displacement (MQD) are obtained by
using the expansion coefficients of $\epsilon$ and $\epsilon^{2}$, respectively.
From the symmetry,
we can rewrite $P(\bm{r};t)$ as $P(\bm{r};t) = P(r;t) / 4 \pi r^{2}$ with $r = |\bm{r}|$.
Also, without loss of generality, we can set the wave number vector $\bm{k}$ parallel to the $z$-direction.
Then we can calculate the Fourier transform in eq~\eqref{displacement_distribution_fourier_laplace} in the spherical coordinates:
\begin{equation}
 \label{displacement_distribution_fourier_laplace_msd_mqd}
 \begin{split}
  \hat{P}(\bm{k};s) 
  & = \int_{0}^{\infty} dt \int_{0}^{\infty} dr \int_{0}^{2\pi} d\theta \int_{0}^{\pi} d\phi \, r^{2} \sin \phi \, e^{- i k r \cos \phi  - s t}
   \frac{P(r;t)}{4 \pi r^{2}} \\
  & = \frac{1}{2} \int_{0}^{\infty} dt \, e^{-s t}
  \int_{0}^{\infty} dr \int_{0}^{\pi} d\phi \, \sin \phi \,
  \left[ 1 + \frac{\epsilon}{2} r^{2} \cos^{2} \phi
  + \frac{\epsilon^{4}}{24} r^{4} \cos^{4} \phi
  \right]
   P(r;t) + O(\epsilon^{3}) \\
%  & = \frac{1}{s} + \int_{0}^{\infty} dt \, e^{-s t}
%  \int_{0}^{\infty} dr \,
%  \left[ \frac{\epsilon}{6} r^{2}
%  + \frac{\epsilon^{4}}{120} r^{4}
%  \right]
%   P(r;t) + O(\epsilon^{3}) \\
  & = \frac{1}{s} + \int_{0}^{\infty} dt \, e^{-s t}
  \left[ \frac{\epsilon}{6} \langle \bm{r}^{2}(t) \rangle
  + \frac{\epsilon^{2}}{120} \langle |\bm{r}(t)|^{4} \rangle
  \right] + O(\epsilon^{3}) .
 \end{split}
\end{equation}
By comparing eqs~\eqref{displacement_distribution_fourier_laplace_expanded}
and \eqref{displacement_distribution_fourier_laplace_msd_mqd}, we can determine
the MSD and the MQD.

%This can be easily shown as follows, by utilizing the symmetry of the system:
%\begin{equation}
% \begin{split}
%  \hat{P}(\bm{k};s) 
%%  & = \int_{0}^{\infty} dt \int d\bm{r} \, e^{- i \bm{k} \cdot \bm{r} - s t}
%%   P(\bm{r};t) \\
%  & = \int_{0}^{\infty} dt \int d\bm{r} \, e^{- s t}
%  \left[ 1 - \frac{1}{2} (\bm{k} \cdot \bm{r})^{2}
%  + \frac{1}{24} (\bm{k} \cdot \bm{r})^{4}
%  \right]
%  P(\bm{r};t) + O(\epsilon^{3}) \\
%  & = 1 +
%  \int_{0}^{\infty} dt \, e^{- s t}
%  \left[ 
%  - \frac{1}{2} k_{\alpha} k_{\beta} \langle r_{\alpha}(t) r_{\beta}(t) \rangle 
%  + \frac{1}{24} k_{\alpha} k_{\beta} k_{\gamma} k_{\delta}
%  \langle r_{\alpha}(t) r_{\beta}(t) r_{\gamma}(t) r_{\delta}(t) \rangle 
%  \right] + O(\epsilon^{3}) \\
%  & = 1 +
%  \int_{0}^{\infty} dt \, e^{- s t}
%  \left[ 
%  \frac{ \epsilon}{6}  \langle \bm{r}^{2}(t)  \rangle
%  + \frac{1}{24} k_{\alpha} k_{\beta} k_{\gamma} k_{\delta}
%  \langle r_{\alpha}(t) r_{\beta}(t) r_{\gamma}(t) r_{\delta}(t) \rangle 
%  \right] + O(\epsilon^{3}) \\
%  & = 1 +
%  \int_{0}^{\infty} dt \, e^{- s t}
%  \left[ 
%  \frac{ \epsilon}{6}  \langle \bm{r}^{2}(t)  \rangle
%  + \frac{1}{24} k_{\alpha}^{4}
%  \langle r_{\alpha}^{4}(t) \rangle 
%  + \frac{1}{8} k_{\alpha}^{2} k_{\beta}^{2}
%  \langle r_{\alpha}^{2}(t) r_{\beta}^{2}(t) \rangle 
%  \right] + O(\epsilon^{3}) \\
%  & =
% \end{split}
%\end{equation}

The coefficient of $\epsilon$ in eq~\eqref{displacement_distribution_fourier_laplace_expanded}
can be calculated as
\begin{equation}
 \label{displacement_distribution_fourier_laplace_1st}
 \begin{split}
  \frac{\tau \hat{\Psi}_{1}(u)}{[1 - \hat{\Psi}_{0}(u)]^{2}}
%  & = \frac{ k_{B} T \tau}{\kappa}
%  \frac{(u + 1)^{2}}{u^{2}}
%  \left( \frac{1}{u + 1} - \frac{1}{u + 1 + \eta} \right) \\
%  & = \frac{ k_{B} T \tau}{\kappa}
%  \frac{u + 1}{u^{2}}\frac{\eta}{u + 1 + \eta} \\
  & = \frac{ k_{B} T \tau}{\kappa}
  \frac{\eta (u + 1)}{u^{2} (u + 1 + \eta)} \\
  & = \frac{ k_{B} T \tau}{\kappa}
\left[ \frac{  \eta }{(1 + \eta) u^{2}}
  + \frac{\eta^{2}}{(1 + \eta)^{2}} \left(\frac{1}{u} - \frac{1}{u + 1 + \eta} \right)\right] .
%  & = \frac{ k_{B} T \tau}{\kappa}
%  \frac{\eta (u + 1)}{u^{2} (u + 1 + \eta)} = \frac{ k_{B} T \tau}{\kappa}
%\left[ \frac{  \phi }{u^{2}}
%  + \phi^{2} \left(\frac{1}{u} - \frac{1}{u + 1 + \eta} \right)\right] ,
 \end{split}
\end{equation}
%where we have defined $\phi \equiv \eta / (1 + \eta)$.
%\begin{equation}
% \begin{split}
%  \frac{\tau \hat{\Psi}_{1}(s)}{[1 - \hat{\Psi}_{0}(s)]^{2}}
%%  & = \frac{\tau}{[1 - \frac{1}{\tau (s + 1 / \tau)}]^{2}} 
%%  \frac{ k_{B} T }{\tau \kappa}
%%  \left( \frac{1}{s + 1 / \tau} - \frac{1}{s + 1 / \tau + \Lambda \kappa} \right) \\
%%  & = \frac{\tau^{3} (s + 1 / \tau)^{2}}{[\tau (s + 1 / \tau) - 1]^{2}} 
%%  \frac{ k_{B} T }{\tau \kappa}
%%  \left( \frac{1}{s + 1 / \tau} - \frac{1}{s + 1 / \tau + \Lambda \kappa} \right) \\
%%  & =
%%  \frac{ k_{B} T }{\kappa}
%%   \frac{(s + 1 / \tau)^{2}}{s^{2}} 
%%  \left( \frac{1}{s + 1 / \tau} - \frac{1}{s + 1 / \tau + \Lambda \kappa} \right) \\
%%  & =
%%  \frac{ k_{B} T }{\kappa}
%%   \frac{s + 1 / \tau}{s^{2}} \frac{\Lambda \kappa}{s + 1 / \tau + \Lambda \kappa} \\
%%  & =
%%  \frac{ k_{B} T }{\kappa}
%%  \left[ \frac{\Lambda \kappa}{1 / \tau + \Lambda \kappa}
%%  \frac{1}{\tau s^{2}}
%%  +  \left(\frac{\Lambda \kappa}{1 / \tau + \Lambda \kappa}\right)^{2}
%%  \left(
%%  \frac{1}{s}
%%  - \frac{1}{s + 1 / \tau + \Lambda \kappa}
%%  \right)
%%  \right] \\
%  & =
%  \frac{ k_{B} T }{\kappa} \frac{\Lambda \kappa}{1 / \tau + \Lambda \kappa}
%  \left[ 
%  \frac{1}{\tau s^{2}}
%  +\frac{\Lambda \kappa}{1 / \tau + \Lambda \kappa}
%  \left(
%  \frac{1}{s}
%  - \frac{1}{s + 1 / \tau + \Lambda \kappa}
%  \right)
%  \right] .
% \end{split}
%\end{equation}
By performing the inverse Laplace transform
for eq~\eqref{displacement_distribution_fourier_laplace_1st}, we have the following expression for the MSD:
\begin{equation}
 \label{msd_explicit_expression_supercooled}
 \langle \bm{r}^{2}(t) \rangle
  = \frac{6 k_{B} T}{\kappa} \frac{  \eta }{1 + \eta}
  \left[  \frac{t}{\tau}
  + \frac{\eta}{1 + \eta} 
  [ 1 - e^{- t (1 + \eta) / \tau}] \right] .
\end{equation}
The coefficient of $\epsilon^{2}$ in
eq~\eqref{displacement_distribution_fourier_laplace_expanded}
can be calculated in a similar way,
although the calculation becomes lengthy:
\begin{equation}
\label{displacement_distribution_fourier_laplace_2nd}
 \begin{split}
  & \frac{\tau \hat{\Psi}_{2}(u)}{[1 - \hat{\Psi}_{0}(u)]^{2}} 
  + \frac{\tau \hat{\Psi}_{1}^{2}(u)}{[1 - \hat{\Psi}_{0}(u)]^{3}} \\
  & = \frac{(k_{B} T)^{2} \tau}{\kappa^{2}}
    \frac{\eta^{2} (u + 1) (u^2 + 2 u + \eta u + 1 + 2 \eta)}{u^{3} (u + 1 + \eta)^{2} (u + 1 + 2 \eta)} \\
%  & = \frac{(k_{B} T)^{2} \tau}{\kappa^{2}}
%  \bigg[
%  - \frac{\eta^{4} (3 + 4 \eta - 2 \eta^{2})}{(1 + \eta)^{4} (1 + 2 \eta)^{2} u}
%  - \frac{\eta^{2} (2 + \eta^{2})}{(1 + \eta)^{4} (u + \eta + 1)}
%  + \frac{\eta^{3}}{(1 + \eta)^{3} (u + 1 + \eta)^{2}} \\
%  & \qquad 
%  + \frac{2 \eta^{2}}{(1 + 2 \eta)^{2} (u + 1 + 2 \eta)}
%  + \frac{\eta^{2}}{(1 + \eta)^{2} u^{3}}
%  + \frac{\eta^{3}( 1 + 3 \eta)}{(1 + \eta)^{3} (1 + 2 \eta) u^{2}}
%  \bigg] \\
  & = \frac{(k_{B} T)^{2} \tau}{\kappa^{2}}
  \bigg[  \frac{\eta^{2}}{(1 + \eta)^{2} u^{3}}
    + \frac{\eta^{3} (1 + 3 \eta)}{(1 + \eta)^{3} (1 + 2 \eta) u^{2}} 
  + \frac{\eta^{3}}{(1 + \eta)^{3} (u + 1 + \eta)^{2}} \\
  & \qquad   + \frac{\eta^{2} (2 + \eta^{2})}{(1 + \eta)^{4}}
  \left( \frac{1}{u}
  - \frac{1}{u + 1 + \eta} \right) 
  - \frac{2 \eta^{2}}{(1 + 2 \eta)^{2} }
  \left( \frac{1}{u}
  - \frac{1}{u + 1 + 2 \eta} \right) \bigg] .
 \end{split}
\end{equation}
The MQD is calculated by performing the inverse Laplace transform of
eq~\eqref{displacement_distribution_fourier_laplace_2nd}:
\begin{equation}
 \begin{split}
 \label{mqd_explicit_expression_supercooled}
  \langle |\bm{r}(t)|^{4} \rangle
%  & = \frac{120 (k_{B} T)^{2}}{\kappa^{2}}
%  \bigg[  \frac{\eta^{2}}{2 (1 + \eta)^{2}} \frac{t^{2}}{\tau^{2}}
%    + \frac{\eta^{3} (1 + 3 \eta)}{(1 + \eta)^{3} (1 + 2 \eta)} 
%  \frac{t}{\tau}
%  + \frac{\eta^{3}}{2 (1 + \eta)^{3}} \frac{t^{2}}{\tau^{2}}
%  e^{-t (1 + \eta) / \tau} \\
%  & \qquad   + \frac{\eta^{2} (2 + \eta^{2})}{(1 + \eta)^{4}}
%  [ 1  - e^{-t (1 + 2 \eta) / \tau} ]
%  - \frac{2 \eta^{2}}{(1 + 2 \eta)^{2} }
%  [ 1 - e^{-t (1 + 2 \eta) / \tau} ]\bigg] \\
  & = \frac{60 (k_{B} T)^{2}}{\kappa^{2}}
  \bigg[  \frac{\eta^{2}}{(1 + \eta)^{2}} \frac{t^{2}}{\tau^{2}}
    + \frac{2 \eta^{3} (1 + 3 \eta)}{(1 + \eta)^{3} (1 + 2 \eta)} 
  \frac{t}{\tau}
  + \frac{\eta^{3}}{(1 + \eta)^{3}} \frac{2 t}{\tau}
  e^{-t (1 + \eta) / \tau} \\
  & \qquad   + \frac{2 \eta^{2} (2 + \eta^{2})}{(1 + \eta)^{4}}
  [ 1  - e^{-t (1 + \eta) / \tau} ]
  - \frac{4 \eta^{2}}{(1 + 2 \eta)^{2} }
  [ 1 - e^{-t (1 + 2 \eta) / \tau} ]\bigg] .
 \end{split}
\end{equation}
Then we can calculate the NGP. From eq~\eqref{msd_explicit_expression_supercooled}, the square of the MSD becomes
\begin{equation}
 \label{msd_explicit_expression_supercooled_squared}
 \begin{split}
  \langle \bm{r}^{2}(t) \rangle^{2}
%  & = \frac{60 (k_{B} T)^{2}}{\kappa^{2}}
%  \left[ \frac{  \eta }{(1 + \eta)} \frac{t}{\tau}
%  + \frac{\eta^{2}}{(1 + \eta)^{2}} 
%  [ 1 - e^{- t (1 + \eta) / \tau}] \right]^{2} \\
  & = \frac{36 (k_{B} T)^{2}}{\kappa^{2}}
  \bigg[ 
  \frac{  \eta^{2} }{(1 + \eta)^{2}} \frac{t^{2}}{\tau^{2}}
  + \frac{2 \eta^{3}}{(1 + \eta)^{3}} 
  \frac{t}{\tau} [ 1 - e^{- t (1 + \eta) / \tau}]  \\
  & \qquad + \frac{\eta^{4}}{(1 + \eta)^{4}} 
  [ 1 - 2 e^{- t (1 + \eta) / \tau}
  + e^{- 2 t (1 + \eta) / \tau}]
  \bigg] .
 \end{split}
\end{equation}
By combining eqs~\eqref{mqd_explicit_expression_supercooled}
and \eqref{msd_explicit_expression_supercooled_squared}, we have
\begin{equation}
 \begin{split}
   \frac{3}{5} \langle |\bm{r}(t)|^{4} \rangle
  - \langle \bm{r}^{2}(t) \rangle^{2} 
  & = \frac{36 (k_{B} T)^{2}}{\kappa^{2}}
  \bigg[ 
  \frac{2 \eta^{4}}{(1 + \eta)^{3} (1 + 2 \eta)} 
  \frac{t}{\tau}
  + \frac{4 \eta^{3}}{(1 + \eta)^{3}} 
  \frac{t}{\tau} e^{- t (1 + \eta) / \tau}
  \\
  & \qquad   + \frac{4 \eta^{2} }{(1 + \eta)^{4}}
  [ 1  - e^{-t (1 + \eta) / \tau} ]
  - \frac{4 \eta^{2}}{(1 + 2 \eta)^{2} }
  [ 1 - e^{-t (1 + 2 \eta) / \tau} ] \\
  & \qquad
  + \frac{\eta^{4}}{(1 + \eta)^{4}} 
  [ 1 - e^{- 2 t (1 + \eta) / \tau}]
   \bigg] .
 \end{split} 
\end{equation}
Finally we have the following explicit expression for the NGP:
\begin{equation}
 \label{ngp_explicit_expression_supercooled}
  \begin{split}
  \alpha(t) 
%   & =     \bigg[ 
%  \frac{2 \eta^{4}}{(1 + \eta)^{3} (1 + 2 \eta)} 
%  \frac{t}{\tau}
%  + \frac{4 \eta^{3}}{(1 + \eta)^{3}} 
%  \frac{t}{\tau} e^{- t (1 + \eta) / \tau}
%  \\
%  & \qquad   + \frac{4 \eta^{2} }{(1 + \eta)^{4}}
%  [ 1  - e^{-t (1 + \eta) / \tau} ]
%  - \frac{4 \eta^{2}}{(1 + 2 \eta)^{2} }
%  [ 1 - e^{-t (1 + 2 \eta) / \tau} ] 
%   + \frac{\eta^{4}}{(1 + \eta)^{4}} 
%  [ 1 - e^{- 2 t (1 + \eta) / \tau}]
%   \bigg] \\
%   & \qquad \times \left[ \frac{  \eta }{(1 + \eta)} \frac{t}{\tau}
%  + \frac{\eta^{2}}{(1 + \eta)^{2}} 
%  [ 1 - e^{- t (1 + \eta) / \tau}] \right]^{-2} \\
   & = \left[ \frac{t}{\tau}
  + \frac{\eta}{1 + \eta} 
  [ 1 - e^{- t (1 + \eta) / \tau}] \right]^{-2}     \bigg[ 
  \frac{2 \eta^{2}}{(1 + \eta) (1 + 2 \eta)} 
  \frac{t}{\tau}
  + \frac{4 \eta}{1 + \eta} 
  \frac{t}{\tau} e^{- t (1 + \eta) / \tau}
  \\
  & \qquad
      + \frac{4 }{(1 + \eta)^{2}}
  [ 1  - e^{-t (1 + \eta) / \tau} ]
- \frac{4 (1 +\eta)^{2}}{(1 + 2 \eta)^{2} }
  [ 1 - e^{-t (1 + 2 \eta) / \tau} ] 
   + \frac{\eta^{2}}{(1 + \eta)^{2}} 
  [ 1 - e^{- 2 t (1 + \eta) / \tau}]
   \bigg] .
  \end{split}
\end{equation}
Eqs~\eqref{msd_explicit_expression_supercooled} and
\eqref{ngp_explicit_expression_supercooled} give
eqs~\eqref{msd_expression_supercooled_final} and
\eqref{ngp_expression_supercooled_final} in the main text.

If the parameter $\eta$ is sufficiently large, two characteristic
time scales ($1 / \Lambda \kappa$ and $\tau$) are
well separated, and both the MSD and the NGP
exhibit several characteristic regions with different $t$ dependence. For the MSD, from eq~\eqref{msd_explicit_expression_supercooled},
we have
\begin{equation}
 \label{msd_explicit_expression_supercooled_approx}
 \langle \bm{r}^{2}(t) \rangle
 \approx \frac{6 k_{B} T}{\kappa}
 \left[  \frac{t}{\tau}
  +  1 - e^{- \Lambda \kappa t} \right] ,
\end{equation}
and thus we find that the MSD exhibits three regions:
\begin{equation}
 \label{msd_explicit_expression_supercooled_approx_regions}
  \frac{\langle \bm{r}^{2}(t) \rangle}{6 k_{B} T / \kappa}
 \approx
 \begin{cases}
   \Lambda \kappa t  & (t \ll 1 / \Lambda \kappa) , \\
  1 & (1 / \Lambda \kappa \ll t \ll \tau) ,\\
  t / \tau & (\tau \ll t) .
 \end{cases}
\end{equation}
%\begin{equation}
% \label{msd_explicit_expression_supercooled_approx_regions}
% \langle \bm{r}^{2}(t) \rangle
% \approx
% \begin{cases}
%  6 k_{B} T \Lambda t  & (t \ll 1 / \Lambda \kappa) , \\
%  6 k_{B} T / \kappa  & (1 / \Lambda \kappa \ll t \ll \tau) ,\\
%  6 k_{B} T t / \kappa \tau & (\tau \ll t) .
% \end{cases}
%\end{equation}
Eq~\eqref{msd_explicit_expression_supercooled_approx} is the same
form as eq~\eqref{msd_supercooled_liquid_scaling}.
For the NGP, from eq~\eqref{ngp_explicit_expression_supercooled}, we
simply have $\alpha(t) \approx \tau / t$
as the approximate form for $\tau \ll t$.
For $1 / \Lambda \kappa \ll t \ll \tau$, we have
\begin{equation}
 \label{ngp_explicit_expression_supercooled_approx_mid}
%  \begin{split}
  \alpha(t) 
%   & = \left[ \frac{t}{\tau}
%  + \frac{\eta}{1 + \eta} 
%  [ 1 - e^{- t (1 + \eta) / \tau}] \right]^{-2}     \bigg[ 
%  \frac{2 \eta^{2}}{(1 + \eta) (1 + 2 \eta)} 
%  \frac{t}{\tau}
%  + \frac{4 \eta}{1 + \eta} 
%  \frac{t}{\tau} e^{- t (1 + \eta) / \tau}
%  \\
%  & \qquad
%      + \frac{4 }{(1 + \eta)^{2}}
%  [ 1  - e^{-t (1 + \eta) / \tau} ]
%- \frac{4 (1 +\eta)^{2}}{(1 + 2 \eta)^{2} }
%  [ 1 - e^{-t (1 + 2 \eta) / \tau} ] 
%   + \frac{\eta^{2}}{(1 + \eta)^{2}} 
%  [ 1 - e^{- 2 t (1 + \eta) / \tau}]
%   \bigg] \\
%   & \approx \left[ \frac{t}{\tau}
%  + \frac{\eta}{1 + \eta}  \right]^{-2}     \bigg[ 
%  \frac{2 \eta^{2}}{(1 + \eta) (1 + 2 \eta)} 
%  \frac{t}{\tau}
%      + \frac{4 }{(1 + \eta)^{2}}
%- \frac{4 (1 +\eta)^{2}}{(1 + 2 \eta)^{2} }
%   + \frac{\eta^{2}}{(1 + \eta)^{2}} 
%   \bigg] \\
    \approx \frac{(1 + \eta)^{2}}{\eta^{2}} \left[ 
  \frac{2 \eta^{2}}{(1 + \eta) (1 + 2 \eta)} 
  \frac{t}{\tau}
      + \frac{4 }{(1 + \eta)^{2}}
- \frac{4 (1 +\eta)^{2}}{(1 + 2 \eta)^{2} }
   + \frac{\eta^{2}}{(1 + \eta)^{2}} 
   \right] \approx \frac{t}{\tau} .
%   & \approx 
%  \frac{t}{\tau}
%   - 3 / \eta \\
%   & \approx
%  \end{split}
\end{equation}
For $t \ll 1 / \Lambda \kappa$, we expand eq~\eqref{ngp_explicit_expression_supercooled}
with respect to $t$ and have
\begin{equation}
 \label{ngp_explicit_expression_supercooled_approx_short}
%  \alpha(t) & = \left[ \frac{t}{\tau}
%  + \frac{\eta}{1 + \eta} 
%  [ 1 - e^{- t (1 + \eta) / \tau}] \right]^{-2}     \bigg[ 
%  \frac{2 \eta^{2}}{(1 + \eta) (1 + 2 \eta)} 
%  \frac{t}{\tau}
%  + \frac{4 \eta}{1 + \eta} 
%  \frac{t}{\tau} e^{- t (1 + \eta) / \tau}
%  \\
%  & \qquad
%      + \frac{4 }{(1 + \eta)^{2}}
%  [ 1  - e^{-t (1 + \eta) / \tau} ]
%- \frac{4 (1 +\eta)^{2}}{(1 + 2 \eta)^{2} }
%  [ 1 - e^{-t (1 + 2 \eta) / \tau} ] 
%   + \frac{\eta^{2}}{(1 + \eta)^{2}} 
%  [ 1 - e^{- 2 t (1 + \eta) / \tau}]
%   \bigg] \\
% & \approx \frac{1}{(1 + \eta)^{2} t^{2}}  \bigg[ 
%  \frac{2 \eta^{2}}{(1 + \eta) (1 + 2 \eta)} 
%  \frac{t}{\tau}
%  + \frac{4 \eta}{1 + \eta} 
%  \frac{t}{\tau} e^{- t (1 + \eta) / \tau}
%  \\
%  & \qquad
%      + \frac{4 }{(1 + \eta)^{2}}
%  [ 1  - e^{-t (1 + \eta) / \tau} ]
%- \frac{4 (1 +\eta)^{2}}{(1 + 2 \eta)^{2} }
%  [ 1 - e^{-t (1 + 2 \eta) / \tau} ] 
%   + \frac{\eta^{2}}{(1 + \eta)^{2}} 
%  [ 1 - e^{- 2 t (1 + \eta) / \tau}]
%   \bigg] \\
  \alpha(t) \approx \frac{1}{(1 + \eta)^{2} (t / \tau)^{2}}  
   \left[ 
 \frac{\eta^{2} (1 + \eta)^{2}}{30} (t / \tau)^{5}
   \right] 
   = \frac{\eta^{2} t^{3}}{30 \tau^{3}} .
%   = \frac{\Lambda^{2} \kappa^{2} t^{3}}{30 \tau} .
\end{equation}
%\begin{equation}
%\begin{split}
% \alpha(t) & = \left[ \frac{t}{\tau}
%  + \frac{\eta}{1 + \eta} 
%  [ 1 - e^{- t (1 + \eta) / \tau}] \right]^{-2}     \bigg[ 
%  \frac{2 \eta^{2}}{(1 + \eta) (1 + 2 \eta)} 
%  \frac{t}{\tau}
%  + \frac{4 \eta}{1 + \eta} 
%  \frac{t}{\tau} e^{- t (1 + \eta) / \tau}
%  \\
%  & \qquad
%      + \frac{4 }{(1 + \eta)^{2}}
%  [ 1  - e^{-t (1 + \eta) / \tau} ]
%- \frac{4 (1 +\eta)^{2}}{(1 + 2 \eta)^{2} }
%  [ 1 - e^{-t (1 + 2 \eta) / \tau} ] 
%   + \frac{\eta^{2}}{(1 + \eta)^{2}} 
%  [ 1 - e^{- 2 t (1 + \eta) / \tau}]
%   \bigg] \\
% & \approx
% \left[ \frac{t}{\tau}
%  + 1 \right]^{-2}     \bigg[ 
%  \frac{2 t}{\tau}
%- 1
%   + 1
%   \bigg] \\
% & \approx 2 t / \tau.
%\end{split}
%\end{equation}
Therefore, we find that the NGP exhibits three regions with different $t$ dependence:
\begin{equation}
 \label{ngp_explicit_expression_supercooled_approx_regions}
  \alpha(t) \approx
  \begin{cases}
   \Lambda^{2} \kappa^{2} t^{3} / 30 \tau & (t \ll 1 / \Lambda \kappa), \\
   \tau / t  & ( 1 / \Lambda \kappa \ll t \ll \tau) , \\
   t / \tau & (\tau \ll t) .
  \end{cases}
\end{equation}
Eqs~\eqref{msd_explicit_expression_supercooled_approx_regions} and
\eqref{ngp_explicit_expression_supercooled_approx_regions} are consistent
with the data in Figure~\ref{letp_msd_and_ngp}.

%------------------------------------------------------------------------------

%\clearpage

%------------------------------------------------------------------------------
% bibliography
\bibliographystyle{apsrev4-1}
\bibliography{coarse_graining_transient_potential}

%\clearpage

%------------------------------------------------------------------------------
% figures

\section* {Figure Captions}

\hspace{-\parindent}%
Figure~\ref{md_msd_and_ngp}: The mean-square displacement (MSD) and non-Gaussianity parameter (NGP)
data of binary Lennard-Jones fluids. The temperatures are set as $k_{B} T = 0.4, 0.5,
0.6, 0.7, 0.8, 0.9$, and $1.0$. For relatively low temperature systems, the
MSD exhibits three characteristic regions, and the NGP becomes large.
See Appendix~\ref{molecular_dynamics_simulation_for_supercooled_liquid} for the
details of the simulations.

\

\hspace{-\parindent}%
Figure~\ref{md_trajectories}: Trajectories of some particles in a supercooled
fluid at $k_{B} T = 0.6$. Points represent the positions of particles at every $1$ unit time scale.
(The size of points is much smaller than the particle size.)
The thick black bars are the scale bars of which
length is the unit length scale $\sigma$.
See Appendix~\ref{molecular_dynamics_simulation_for_supercooled_liquid} for the
details of the simulations.

\

\hspace{-\parindent}%
Figure~\ref{letp_msd_and_ngp}: (a) The mean-square displacement (MSD)
and (b) the non-Gaussianity parameter (NGP) of a particle in a supercooled
liquid by the LETP model, with various average waiting times $\tau$.
The time $t$ is normalized by the characteristic time scale of the 
motion in the transient potential, $1 / \Lambda \kappa$. Also, the
MSD is normalized by the characteristic length of the transient potential.

%\clearpage

%------------------------------------------------------------------------------
\section* {Figures}

%\clearpage

%\clearpage

%------------------------------------------------------------------------------
\end{document}